\documentclass{aa}  
\usepackage{amsmath}
\allowdisplaybreaks
\usepackage{txfonts}
\usepackage{graphicx}
\usepackage{lipsum}
\usepackage{subcaption}
\usepackage{booktabs,tabularx}
\usepackage{lscape}
\usepackage{float}
\usepackage{placeins}   
\usepackage{natbib}       
\usepackage{array}   
\usepackage{hyperref}
\usepackage{afterpage}
\begin{document}

\title{Cluster Ages to Reconstruct the Milky Way Assembly (CARMA)}

   \subtitle{VI. The age-metallicity relation of in-situ clusters and the age of the Universe}

   \author{Chiara Zerbinati\inst{1,2}\fnmsep\thanks{Corresponding author: chiara.zerbinati2@unibo.it}
        \and Davide Massari\inst{1}
        \and Fernando Aguado-Agelet\inst{3,4}
        \and Cristiano Fanelli\inst{1}
        \and Massimiliano Matteuzzi \inst{1,2}
        \and Santi Cassisi\inst{5,6}
        \and Matteo Monelli\inst{5,7,4}
        \and Cristiana Spingola\inst{9}
        \and Carme Gallart\inst{7,4}
        \and Maurizio Salaris\inst{1,10}
        \and Roger E. Cohen \inst{11,12}
        \and Edoardo Ceccarelli\inst{13,1,2}
        \and Sara Saracino\inst{14,10}
        \and Elena Pancino\inst{14}
        }

   \institute{Osservatorio di Astrofisica e Scienza dello Spazio di Bologna, INAF, Via Gobetti 93/3, Bologna, I-40129, Italy
   \and Dipartimento di Fisica e Astronomia, Università degli Studi di Bologna, Via Piero Gobetti 93/2, Bologna, I-40129, Italy
   \and Escola de Enxeñar\'ia de Telecomunicaci\'on, Universidade de Vigo, Vigo, 36310, Spain
   \and Universidad de La Laguna, Avda. Astrof\'isico Fco. S\'anchez, La Laguna, E-38205, Tenerife, Spain
   \and Osservatorio Astronomico di Abruzzo, INAF, Via M. Maggini, Teramo, 64100, Italy
   \and INFN - Sezione di Pisa, Università di Pisa, Largo Pontecorvo 3, Pisa, 56127, Italy
   \and Instituto de Astrof\'isica de Canarias, Calle V\'ia L\'actea s/n, E-38206 La Laguna, Tenerife, Spain
   \and INAF-Osservatorio Astronomico di Roma, Via Frascati 33, 00078 Monte Porzio Catone, Italy
   \and INAF - Istituto di Radioastronomia, via P. Gobetti 101, 40129, Bologna, Italy
   \and Astrophysics Research Institute, Liverpool John Moores University, 146 Brownlow Hill, Liverpool L3 5RF, UK
   \and Department of Physics and Astronomy, Rutgers the State University of New Jersey, 136 Frelinghuysen Road, Piscataway, 08854, NJ, USA
   \and Eureka Scientific, 2452 Delmer St., Suite 100, Oakland, CA 94602, USA
   \and Kapteyn Astronomical Institute, University of Groningen, Landleven 12, 9747 AD Groningen, The Netherlands
   \and INAF - Osservatorio Astrofisico di Arcetri, Largo E. Fermi 5, I-50125 Firenze, Italy}

\abstract{Within the homogeneous framework for the derivation of globular clusters ages defined by the CARMA collaboration, we focus on the clusters born {\it in-situ} in the Milky Way (MW). As they are the oldest among the old globular clusters, their absolute age provides crucial constraints about the age of the Universe. Moreover, the shape of their age-metallicity relation (AMR) is expected to show the imprints of past merger events, in case these were able to trigger cluster formation in the Galaxy.
We derive new homogeneous ages for 18 GCs dynamically and/or chemically classified as in-situ, and build the MW AMR from a total sample of 34 GCs. We find that all the selected GCs follow a well defined AMR, with the exception of NGC 6642 and Pal 1. While the former is likely a metal-rich and low-orbital-energy member of the LKH merger event, and thus puts the chronological evidence for the existence of this progenitor on even more robust ground, for the latter we propose an alternative scenario according to which it formed in-situ, but in a metal-depleted region of the Galaxy, possibly due to the interaction with an unspecified merger event.
To find evidence of episodic, merger-driven GC formation, we use a Gaussian Mixture Model to describe the observed age distribution. This is statistically best-fit by 4 components, which we associate with significant events of the Milky Way merger history.
Cluster ages were finally calibrated to a more accurate absolute age scale based on asteroseismic measurements of the mass of red giant stars in NGC 6791, and the resulting AMR was fit to compute the look-back time for the beginning of cluster formation in the Milky Way. The fit provides $t_{GC,i}=13.59^{+0.33}_{-0.32}$ Gyr for the initial time of cluster formation, and $t_{U}=13.79\pm 0.34$ Gyr for the age of the Universe. This value translates to $H_{0}=68.72^{+3.95}_{-3.49}\; \text{(stat.)} ^{+2.65}_{-2.32}\; \text{(sys.)} \; \text{km s}^{-1} \text{Mpc}^{-1},$ where the systematic error comes from a realistic estimate of the zero point uncertainty on the GC absolute ages derived from stellar models. This result is in good agreement with probes from the early Universe, though still consistent with local probes within the uncertainties.}
\maketitle
\nolinenumbers
\section{Introduction}
According to the $\Lambda$-Cold Dark Matter ($\Lambda$CDM) cosmological paradigm, the Milky Way (MW) evolved following a hierarchical merging scenario, building up its mass and current structure from the accretion of smaller galaxies. In the last decades, this formation path acquired many pieces of evidence in its support, the first being the discovery of the ongoing merger event with the Sagittarius dwarf galaxy \citep{1994Ibata}.
With the availability of precise six-dimensional phase space information provided by the ESA-Gaia mission \citep{2016Gaia,2023_GDR3}, and with its combination with chemical abundances obtained by means of the most recent spectroscopic surveys, e.g. APOGEE \citep{2017APOGEE}, GALAH \citep{2015galah}, the Gaia ESO Survey \citep{2012gaiaeso}, and LAMOST \citep{2012lamost}, a number of past merger events have been identified and characterized \citep[see][for a review]{Helmi_review}. \\
The most recent discoveries reveal that the MW halo hosts the debris of at least two significant merger events, named LKH \citep{Massari_2019,2019Kraken,Horta_2020,massari26} and Gaia-Sausage-Enceladus \citep[GSE, see][]{2018NatureGSE,Belokurov_2018}, respectively, as well as the relics of less massive galaxies, e.g. Sequoia \citep{Myeong_2019}, the Helmi streams \citep{1999Hstreams}, Antaeus \citep{Oria_2022,ceccarelli2024walkretrogradewrsproject} and other smaller substructures \citep[see][]{Koppelman_2019,Horta_2020,Malhan_2022,RLAra_2022,Mikkola_2022,Dodd_2023}.   \\ 
In this context, globular clusters (GCs) constitute ideal tracers to investigate the MW assembly history, as they are among the oldest objects we can observe in the Galaxy. To a good approximation, they can be treated as single stellar populations, with stars all born at the same time and with similar chemical composition, at least in certain chemical elements \citep{2004Gratton}. Therefore, they are expected to have retained both chemical and dynamical information about the environment in which they were born.\\
The dynamics of GCs has proved to be effective in disentangling between different merger events \citep[e.g. in][]{Massari_2019,forbes2020,Callingham_2022,De_Leo_2026}. However, it has become clear that the evolution processes experienced by the Milky Way have at least partially erased the dynamical coherence of GCs \citep[e.g.,][]{2023Pagnini,2024Chen,2026Istiak}. Additional information is therefore required to appropriately reconstruct the Milky Way assembly with GCs. In this sense, a crucial piece of information is provided by GC ages. Indeed, a powerful tool to distinguish between different accretion events is the age-metallicity relation (AMR) of the associated stellar populations. Already since the works of \cite{2009MF} and \cite{2010FB}, evidence was found that the AMR of MW GCs bifurcates in the metal-rich ([Fe/H]$\gtrsim$-1.3) regime, separating accreted from in-situ systems. However, the efficacy of the AMR in distinguishing accreted from in-situ clusters in the metal-poor regime, as well as GCs born in different accreted progenitors, has been hampered so far by large systematic uncertainties in the methods, models and observations. Homogeneous estimates significantly improved the age precision \citep[e.g][]{VandenBerg_2013}, but have lacked detailed treatment of each GC colour-magnitude diagram, including differential reddening correction or proper-motion-based field decontamination until recently \citep{Massari_2023,valcin2026ageuniverseglobularclusters}. \\
The CARMA project \citep{Massari_2023}, aims at overcoming these limitations, by building the most precise (with uncertainties <0.5 Gyr) and complete catalogue of MW GC ages to date. This is achieved by strictly analysing high-resolution Hubble Space Telescope photometry in F606W and F814W bands, and by adopting the most up-to-date set of BaSTI theoretical isochrones \citep{2018Hidalgo,10.1093/mnras/stad3267}.\\
The method turned out to be successful in solving cases of debated GC origin \citep{Massari_2023,ceccarelli2025clusteragesreconstructmilky}, in characterizing individual merger events like GSE \citep{Aguado_Agelet_2025} or in investigating the assembly history of smaller galaxies like the LMC \citep{niederhofer2025clusteragesreconstructmilky}, and also in identifying the GCs that were brought in the MW by the very ancient merger dubbed LKH \citep[][M26 hereafter]{massari26}, which was characterized in terms of mass and accretion time, too.\\ 
In this sixth paper of the CARMA series, we focus on the AMR of the Milky Way in-situ GCs. On the one hand, in-situ GCs provide an empirical tool to set constraints on the initial time for the beginning of star formation and for the age of the Universe independently of cosmological assumptions \citep[see][]{cimatti2023revisitingoldeststarscosmological}, given that they are among the oldest stellar populations we can observe. 
It follows that determining the age of the Universe with this method offers a way to estimate the Hubble constant $H_0$ based on GC ages. Such an approach could provide pivotal insights into solving the so called "Hubble tension", i.e. the persistent discrepancy between the $H_0$ measurements derived from early Universe probes at high redshift and those obtained from local observations \citep[see][for a review]{2024_Verde}.    
This underscores why the pursuit of accurate measurements of stellar ages has become a major focus in astrophysics \citep[see e.g.][]{Valcin_2020,valcin2026ageuniverseglobularclusters}. In the past few years, asteroseismic studies took giant steps in establishing a more accurate absolute age scale for stars and GCs, mainly thanks to the determination of masses and radii, for which the estimates are independent of reddening and distance \citep[see][]{Brogaard_2011}.\\
On the other hand, the AMR of in-situ GCs is predicted by cosmological simulations \citep[e.g.][]{2019Krui,2020Hughes,2026Sante} and confirmed by observations \citep[e.g.][]{2019Gallart,2025_Fern_GSE} to be an efficient tool to reconstruct the timeline of the past merger events experienced by the Milky Way, which may have triggered events of in-situ star - and cluster - formation.\\
In this work, in Section \ref{sec:refin} we refine and characterize the MW GC AMR by expanding our catalogue of relative ages for in-situ GCs, and in Section \ref{sec:absolute} we use asteroseismic estimates for the mass of RGB stars in NGC 6791 to calibrate them on a more accurate absolute age scale. In Section \ref{sec:h0} we fit the absolute AMR to derive a value for the age of the Universe and the Hubble constant $H_0.$ Finally, in Section\ref{sec:adf} we investigate the age distribution of the in-situ GCs in relation to the MW assembly history. \\
\section{Data and methodology} \label{sec:meth}
Given that the main requirement of the CARMA project is homogeneity, the data and methods used in our analysis follow those adopted in \cite{Massari_2023}, which we will refer to as Paper I hereafter.
\subsection{The globular cluster data set} \label{sec:dataset}
Our sample includes 18 GCs, dynamically associated with the Milky Way according to \cite{Massari_2019}, eDR3 edition\footnote{see \citep{Massari_2025} and the online table at \url{https://www.oas.inaf.it/en/research/m2-en/carma-en/}}. The complete data set is listed in Table \ref{tab:iso}, together with the values of [M/H], E(B-V), distance modulus, and age derived in our analysis.\\
For the entire set, we used Hubble Space Telescope (HST) photometry, taken with the Advanced Camera for Survey (ACS) in the F606W and F814W optical filters.\\
The photometry for 12 of our GCs was taken from the HST UV Globular Cluster Survey (HUGS) project \citep{Hugs}. We selected the catalogues labelled "method 2", that are obtained with a photometric method which is suitable for faint stars and crowded environments.\\
For the remaining six GCs (namely NGC 6325, NGC 6342, NGC 6355, NGC 6380, NGC 6558 and NGC 6642) the photometry was obtained by \cite{2021_Cohen}.\\
The HUGS catalogue provides membership probabilities for stars in each GC, and given that our analysis benefits more from purity than from completeness, we selected only stars having a probability of membership $>90\%$.\\
Following the same criterion, for the six GCs taken from the Cohen catalogue we performed a proper motion decontamination by selecting as member stars those having a distance smaller than 0.3-0.5 mas/yr with respect to the mean motion of the cluster.\\
After the membership selection, a differential reddening correction was applied, based on the method illustrated in \cite{Milone12}. 
\subsection{Age Derivation} \label{sec:agest}
Ages have been derived following the isochrone fitting procedure described in detail in Paper I.\\
In short, we built a grid of theoretical isochrones from the latest BaSTI database \citep[see][]{10.1093/mnras/stad3267}, spanning an age range going from 6 up to 15 Gyr in bins of 100 Myr, and a metallicity range from -2.5 to +0.3 dex, in steps of 0.01 dex.\\
We selected solar-scaled models accounting for diffusive processes, given that in the F606W and F814W filters, the effect of the specific $\alpha-$elements distribution on the bolometric corrections become negligible, as shown in \cite{Cassisi_2004}. As a consequence, two isochrones at fixed age, with different $[\alpha/Fe]$ but same global metallicity are equivalent in the optical filters we used. Therefore, we can avoid making any assumption on the individual $\alpha-$element abundances and we can work with global metallicities, rather than with iron abundances. The two quantities can indeed be related through the equation \citep[][]{1993Salaris,Cassisi_2004}:
\begin{equation} \label{eq:salaris}
    [M/H]=[Fe/H] + \log(0.694\times 10^{[\alpha/Fe]}+0.306).
\end{equation}
The values of [$\alpha$/Fe] were only estimated after the isochrone fit, in order to compare the results with available spectroscopic [Fe/H] measurements. Specifically, [$\alpha$/Fe] were estimated by using the observed relation between [$\alpha$/Fe] vs [Fe/H] for in-situ stars in the APOGEE catalogue, which is depicted as:
\begin{equation}
\left[ \frac{\alpha}{\text{Fe}} \right] = 
\begin{cases} 
0.4 - \left( \left[ \frac{\text{Fe}}{\text{H}} \right] + 2 \right) \times 0.1 & \text{for } -2.5 < \left[ \frac{\text{Fe}}{\text{H}} \right] < -0.5 \\[3ex]
0.25 - \left( \left[ \frac{\text{Fe}}{\text{H}} \right] + 0.5 \right) \times 0.4 & \text{for } -0.5 < \left[ \frac{\text{Fe}}{\text{H}} \right] < 0.
\end{cases}
\end{equation}
Our isochrone fitting procedure also takes into account temperature-dependent extinction, by correcting the shape of each isochrone according to the prescription by \cite{2008Girardi}.\\
For the isochrone fitting code, we fixed Gaussian priors on [M/H] and E(B-V) based on the Harris catalogue \citep{harris96}, while for the distance moduli (DM) we relied on the estimates in \cite{2021Baumgardt}. \\
The typical sigma values associated with our Gaussian priors are $\sigma_{[M/H]}=0.1$ dex for the metallicity, $\sigma_{E(B-V)}=0.05$ mag for the colour excess, and $\sigma_{DM}=0.1$ mag for the distance modulus. These were eventually adjusted to slightly smaller values only in cases where our best fit converged to solutions that were too far away from the spectroscopic metallicities or the distance moduli, which might happen for the most extincted cases, where the photometry is worse and the differential reddening residual larger.\\
Once the priors are established, the code estimates the goodness of fit by means of a likelihood function made up of two terms: the first term quantifies the distance of the solution from the adopted priors, while the second term takes into account the distance of each individual star in the CMD from the theoretical model.\\
The fitting is done through a Markov-Chain Monte Carlo (MCMC) sampling, and we performed the procedure in both the $(m_{F606W}-m_{F814W})$ vs $m_{F606W}$ and the $(m_{F606W}-m_{F814W})$ vs $m_{F814W}$ CMDs. We selected only stars around the observed mean ridge line of the CMD to avoid the inclusion of binaries and outliers.\\
The results were taken as the average values between the two solutions, whereas the final errors encompass the uncertainties of the two runs combined. 
\begin{table}[h!]
\caption{Results of isochrone fitting: for each cluster, values of [M/H], $E(B-V),$ distance modulus and age are reported along with their associated upper and lower limits.}
\label{tab:iso}
\renewcommand\arraystretch{1.5}
\[
\begin{array}{ccccc} 
\toprule
\hline
\text{Cluster Name} & \text{[M/H]} & \text{E(B-V)}  & \text{DM} & \text{Age}\\
                   &              &   \text{[mag]}  &  \text{[mag]} & \text{[Gyr]}\\
\midrule
\text{NGC 104} & -0.60^{+0.01}_{-0.01} & 0.04^{+0.01}_{-0.01} & 13.26^{+0.01}_{-0.01} & 12.84^{+0.12}_{-0.20}\\
\text{Lynga7} & -0.47^{+0.04}_{-0.04} & 0.73^{+0.01}_{-0.01} & 14.45^{+0.01}_{-0.01} & 11.95^{+0.21}_{-0.24}\\
\text{NGC 6171} & -0.84^{+0.03}_{-0.03} & 0.42^{+0.01}_{-0.01} & 13.68^{+0.01}_{-0.01} & 13.46^{+0.27}_{-0.23}\\
\text{NGC 6325} & -1.21^{+0.04}_{-0.04} & 0.95^{+0.01}_{-0.01} & 14.38^{+0.02}_{-0.02} & 13.75^{+0.28}_{-0.20}\\
\text{NGC 6342} & -0.61^{+0.01}_{-0.01} & 0.50^{+0.01}_{-0.01} & 14.74^{+0.02}_{-0.02} & 12.93^{+0.49}_{-0.52}\\
\text{NGC 6355} & -1.17^{+0.1}_{-0.02} & 0.81^{+0.01}_{-0.01} & 14.48^{+0.01}_{-0.01} & 13.51^{+0.35}_{-0.47}\\
\text{NGC 6366} & -0.45^{+0.05}_{-0.06} &  0.71^{+0.01}_{-0.01} & 12.67^{+0.01}_{-0.02} & 11.98^{+0.28}_{-0.21}\\
\text{NGC 6380} & -0.55^{+0.09}_{-0.09} &  1.25^{+0.01}_{-0.01} & 14.78^{+0.02}_{-0.01} & 13.09^{+0.21}_{-0.43}\\
\text{NGC 6558} & -1.00^{+0.01}_{-0.02} & 0.36^{+0.01}_{-0.01} & 14.63^{+0.04}_{-0.03} & 14.29^{+0.26}_{-0.47}\\
\text{NGC 6624} & -0.48^{+0.02}_{-0.02} & 0.26^{+0.01}_{-0.01} & 14.59^{+0.01}_{-0.01} & 12.67^{+0.14}_{-0.12}\\
\text{NGC 6637} & -0.49^{+0.01}_{-0.01} & 0.16^{+0.01}_{-0.01} & 14.69^{+0.01}_{-0.01} & 13.84^{+0.06}_{-0.08}\\
\text{NGC 6642} & -0.92^{+0.01}_{-0.08} & 0.42^{+0.01}_{-0.01} & 14.54^{+0.02}_{-0.02} & 12.14^{+0.27}_{-0.06}\\
\text{NGC 6652} & -0.60^{+0.01}_{-0.01} & 0.12^{+0.01}_{-0.01} & 14.88^{+0.01}_{-0.01} & 12.93^{+0.22}_{-0.10}\\
\text{NGC 6717} & -1.11^{+0.03}_{-0.05} & 0.21^{+0.01}_{-0.01} & 14.35^{+0.02}_{-0.02} & 13.82^{+0.24}_{-0.80}\\
\text{NGC 6723} & -0.83^{+0.02}_{-0.01} & 0.07^{+0.01}_{-0.01} & 14.56^{+0.01}_{-0.01} & 13.11^{+0.13}_{-0.11}\\
\text{NGC 6791} & +0.34^{+0.03}_{-0.02} & 0.15^{+0.01}_{-0.01} & 13.12^{+0.01}_{-0.01} & 9.15^{+0.23}_{-0.10}\\
\text{NGC 6838} & -0.58^{+0.02}_{-0.01} & 0.23^{+0.01}_{-0.01} & 13.00^{+0.01}_{-0.01} & 12.80^{+0.16}_{-0.15}\\
\text{Pal1} & -0.46^{+0.04}_{-0.04} & 0.13^{+0.01}_{-0.01} & 15.23^{+0.02}_{-0.03} & 9.40^{+0.28}_{-0.39}\\
\bottomrule
\end{array}
\]
\end{table}

\afterpage{
 \begin{figure*}[pt]
   \centering
   \begin{minipage}{0.45\textwidth}
        \centering
        \includegraphics[width=1.0\textwidth]{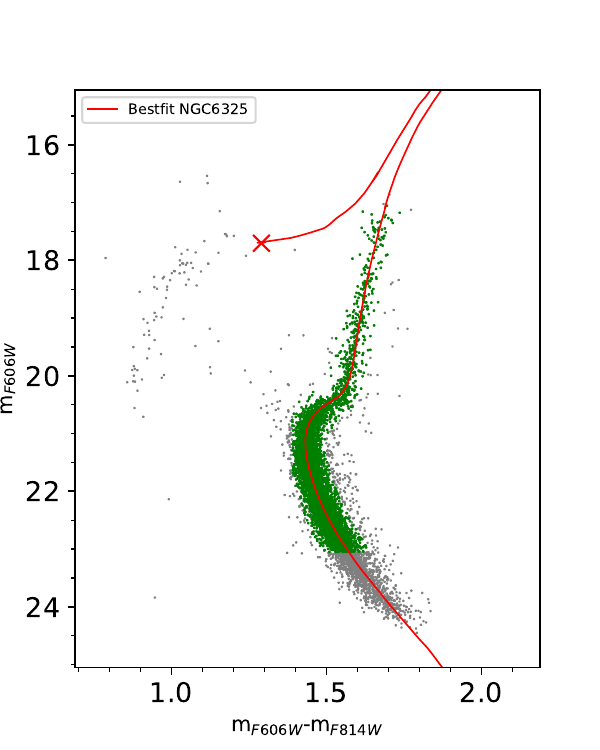}
        \includegraphics[width=1.0\textwidth]{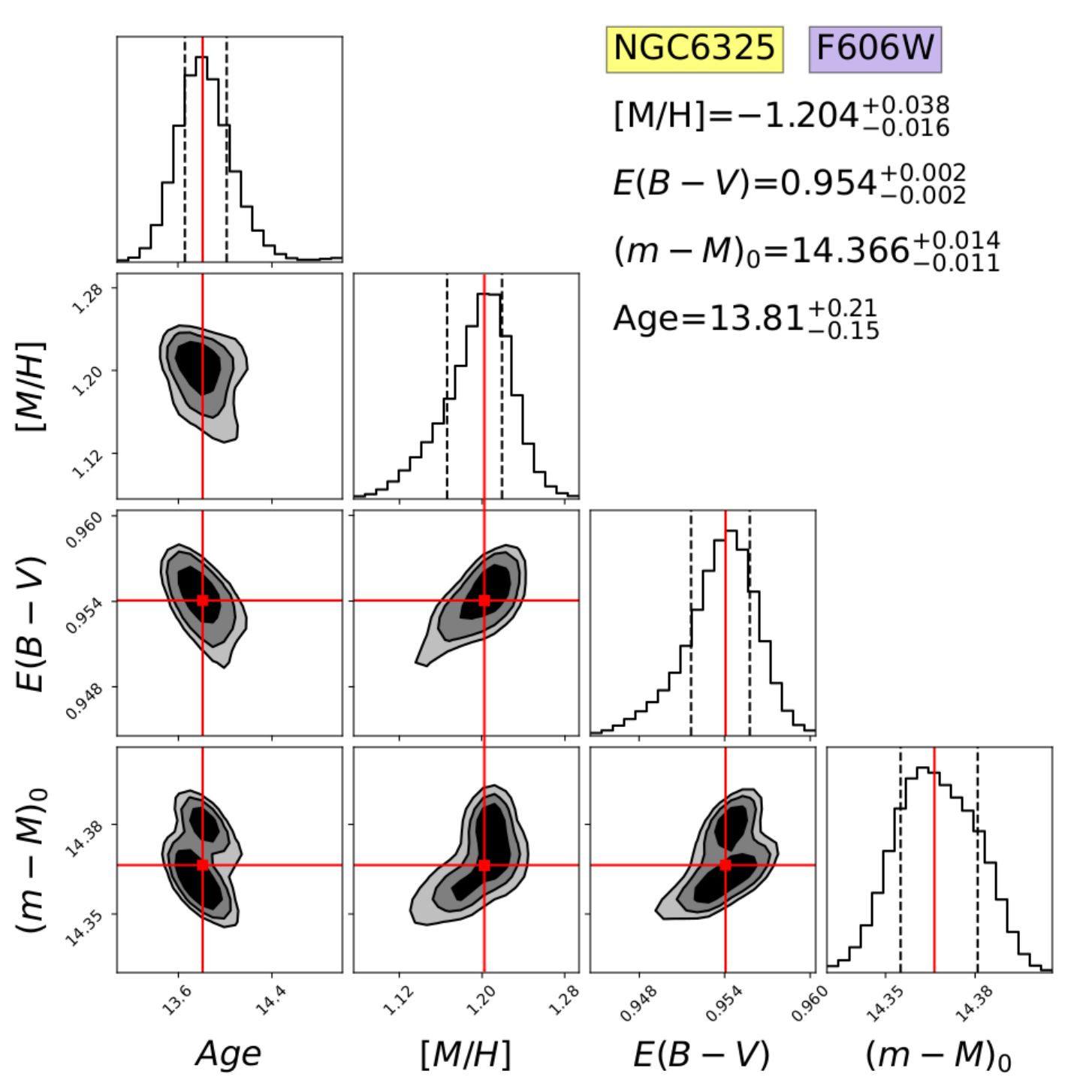}
   \end{minipage}\hfill
   \begin{minipage}{0.45\textwidth}
        \centering
        \includegraphics[width=1.0\textwidth]{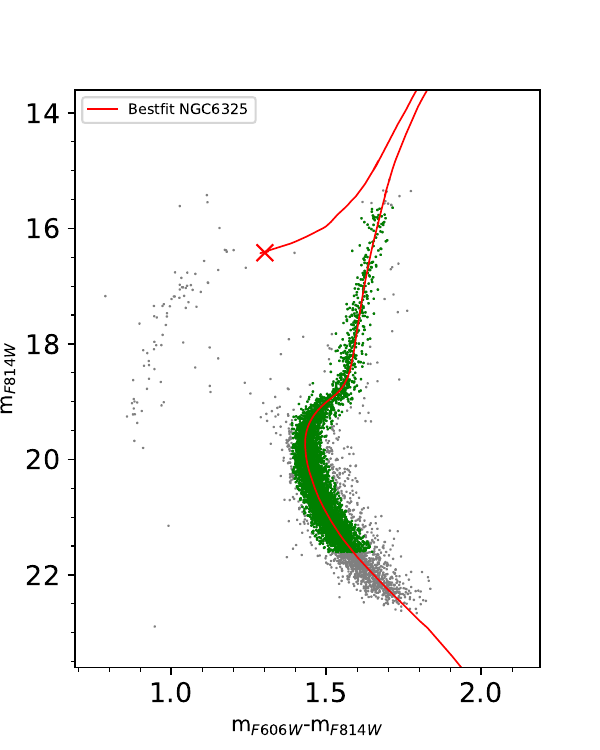}
        \includegraphics[width=1.0\textwidth]{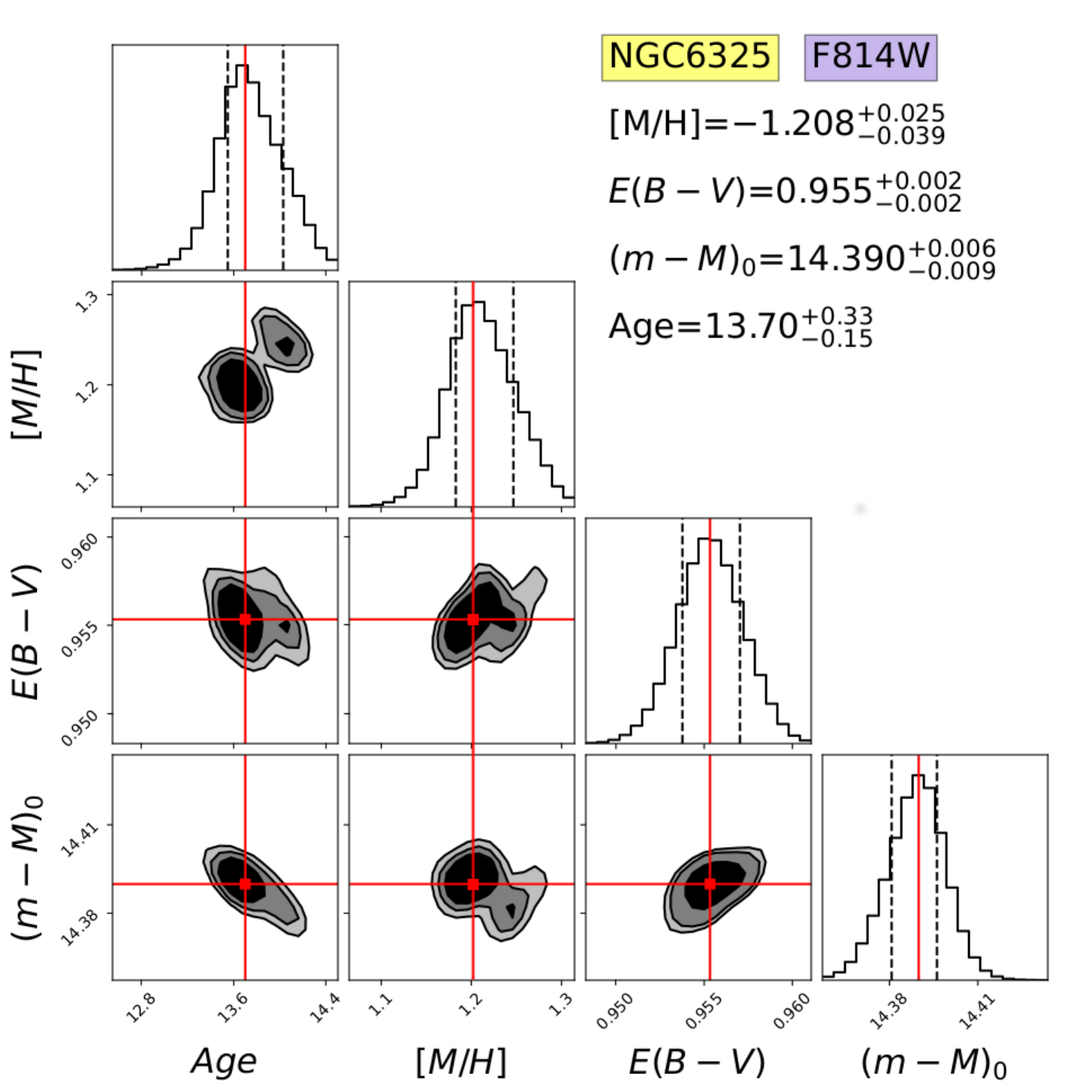}
   \end{minipage}
   \caption{{Results of the isochrone fit for NGC~6325.} Top left panel: best fit model in the ($\it{m_{\mathrm{F814W}}}$, $\it{m_{\mathrm{F606W}}} - \it{m_{\mathrm{F814W}}}$) CMD. Top right panel: best fit model in the ($\it{m_{\mathrm{F606W}}}$, $\it{m_{\mathrm{F606W}}} - \it{m_{\mathrm{F814W}}}$) CMD. Bottom left panel: posterior distributions for the output parameters and the best-fit solution, quoted in the labels, in the ($\it{m_{\mathrm{F814W}}}$, $\it{m_{\mathrm{F606W}}} - \it{m_{\mathrm{F814W}}}$) CMD. Bottom right panel: posterior distributions for the output parameters and the best-fit solution, quoted in the labels, in the ($\it{m_{\mathrm{F606W}}}$, $\it{m_{\mathrm{F606W}}} - \it{m_{\mathrm{F814W}}}$) CMD.}
              \label{fig:ages_6325}
    \end{figure*}
    \clearpage}

\subsection{AMR models} \label{sec:model}
Following the derivation of age estimates, we modelled the resulting age-metallicity relation with the aim of constraining the onset of GC formation in the Milky Way. \\ 
We tested two possibilities for the evolution of the Star Formation Rate (SFR) in the MW: the case with constant SFR, and the one with exponentially declining SFR (see M26 for details about their derivation).\\ 
The former represents a leaky-box model \citep{Prantzos_2008} where the effective yield $p$ has been derived empirically in \cite{2003Dekkel} by fitting 40 galaxies of the Local Group. This model can be considered as representative of Milky-Way-like galaxies in the earliest stages of their evolution, and is expressed through:
\begin{equation} \label{eq:const_SFR}
    t(x)=t_{f}+ \Delta t \cdot \exp \left[\frac{-10^x}{p}\right],
\end{equation}
where $x=[M/H]$ is the metallicity of the system, $p$ is the effective chemical yield (in units of $Z_{\odot}$), and $\Delta t= t_{i}-t_{f}$ is the difference between the time $t_{i}$ at which star formation starts, and the time $t_{f}$ at which it ends. The timescales were measured in Gyr as look-back times, therefore it must hold $t_{i}>t_{f}$ \citep[see e.g.][]{Prantzos_2008,Leaman_2013}.\\
In the second model, the MW SFR is described as an exponentially declining function, to represent the quenching that can be induced by massive mergers such as Gaia-Sausage-Enceladus \citep[][]{2018NaturHelmi,Belokurov_2018}. In this case, the SFR follows $\varpropto \exp (-\beta\:t),$ and the relation between age and metallicity becomes:
\begin{equation} \label{eq:decl_SFR}
    t(x)=t_{i}- 10^{x}/(\beta \: p),
\end{equation}
where in this case $t_{i}\gg t_{f}.$ If this condition does not hold, Eq. \ref{eq:decl_SFR} reduces to Eq. \ref{eq:const_SFR}.
\begin{figure}[t]
\includegraphics[width=1\linewidth]{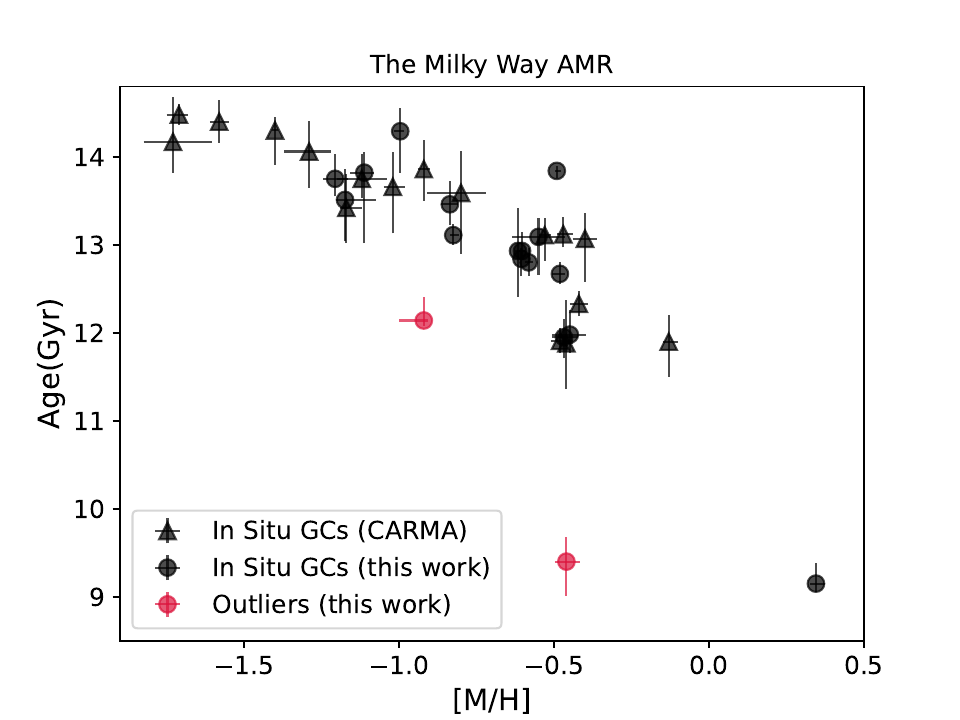}
\caption{Distribution in the age-metallicity plane of the GCs analysed in this work (filled circles) and of the in-situ ones identified in the previous CARMA papers (black triangles).}
\label{fig:amr}
\end{figure}
\quad
\begin{figure}[t]
\includegraphics[width=1\linewidth]{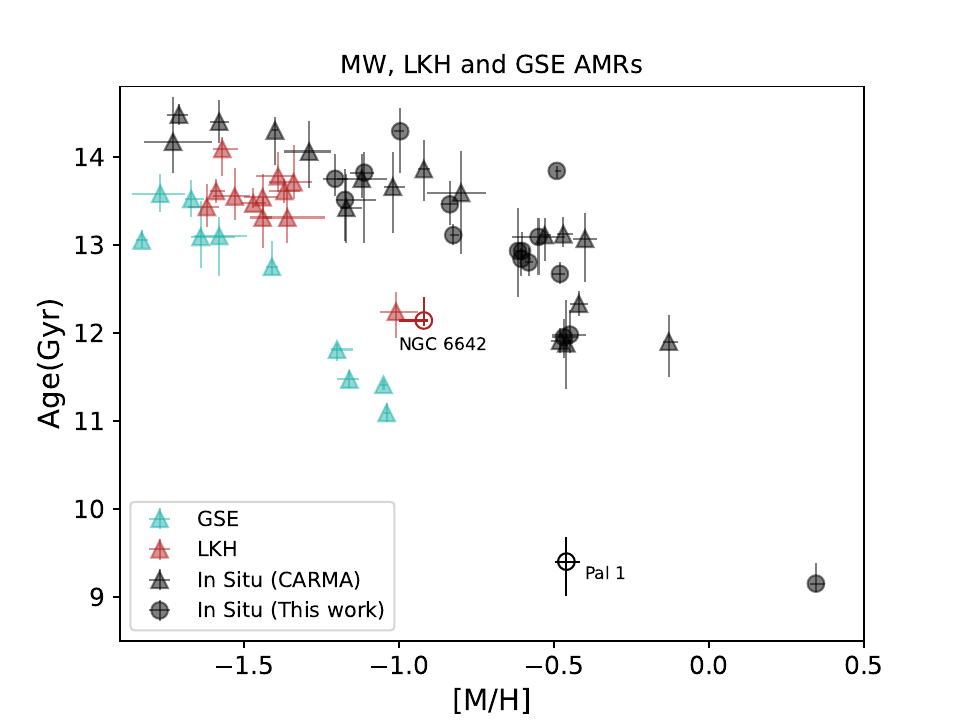}
\caption{Comparison between the location in the age-metallicity plane of in-situ (black triangles), LKH (red triangles) and GSE (cyan triangles) GCs from M26 and the GCs analysed here (filled circles). NGC 6642 and Pal 1, which are off of the track followed by in-situ GCs are highlighted with labels. }
\label{fig:amrtot}
\end{figure}
\section{Results} \label{sec:results}
The results of the isochrone fitting for the entire set of GCs are summarized in Table \ref{tab:iso}. We report an example of the isochrone fitting outputs in Fig. \ref{fig:ages_6325} for NGC 6325, along with the associated corner plots from the MCMC run.\\ 
The plots in the upper panel display the best-fit isochrones in the $m_{F606W}$ vs ($m_{F606W}-m_{F814W}$) and $m_{F814W}$ vs ($m_{F606W}-m_{F814W}$) CMDs on the left and right side, respectively. Stars selected for the fitting process (based on the criterion mentioned in \ref{sec:agest}) are coloured in green. \\The associated corner plots are reported in the lower panel, displaying the MCMC posterior distributions for [M/H], E(B-V), distance modulus and age with the associated uncertainties. \\
Our GCs span a range in metallicities from [M/H]=-1.75 up to [M/H]=+0.34, and a range of ages from 9.15 up to 14.29, with NGC 6791 being the youngest star cluster, and NGC 6558 being the oldest GC.\\
We combined our set of Milky Way GCs with those
already classified as in-situ in previous CARMA papers \citep[see Paper I,][]{Aguado_Agelet_2025,ceccarelli2025clusteragesreconstructmilky,Massari_2025,garro25}, thus increasing the total sample of in-situ clusters to 34. The resulting AMR is shown in Fig. \ref{fig:amr}.\\
Appendix \ref{app:comparisons} reports the comparison between the isochrone fitting outputs and the values expected from the literature for [M/H], E(B-V), DM and ages, respectively. For the metallicity and colour excess, our results are compared to those from the Harris catalogue; distance moduli are compared to the \cite{2021Baumgardt} catalogue, while the ages are compared on a relative scale to those found by \cite{VandenBerg_2013} (labelled in purple) and \cite{2021_Cohen} (labelled in yellow). For all the explored parameters, the comparison with the literature is remarkably good. The average differences are well consistent with zero, and a few outliers are caused by clusters with high extinction. In fact, most of Harris E(B-V) estimates come from isochrone fit, that provide distance and reddening altogether, whereas we prefer to use \cite{2021Baumgardt} as a reference for distance. This means that there exist clusters with DM in good agreement with the Baumgardt compilation, but more discrepant compared to Harris, and hence with its reddening values. 
\subsection{Refinement of the GC associations} \label{sec:refin}
As it can be appreciated in Fig. \ref{fig:amr}, the majority of GCs in our data set follow a well defined AMR, consistent with that defined by the in-situ GCs in M26 (shown as black triangles in Fig. \ref{fig:amrtot}), while two GCs - namely NGC 6642 and Palomar 1 (Pal 1) - look like clear outliers.\\
To better understand their origin, in Fig. \ref{fig:amrtot}, we adapted the AMR in M26 adding our GC set, colour-coded according to the new associations. From this comparison, NGC 6642 appears too young to be considered of in-situ origin, and its location in the AMR is instead well consistent with being a member of the accreted LKH progenitor (red triangles). Such an interpretation holds even when considering NGC 6642 orbital parameters. In fact, according to its Jacobi energy ($E_J$), its vertical angular momentum ($L_z$) and its circularity obtained as in \cite{Massari_2019}, this GC appears coherent with the LKH population as well. \\
The discovery of another metal-rich GC associated with LKH puts the chronological evidence for the existence of this progenitor on even more solid ground. In a recent study, \cite{laporte2026galacticfireworktimingsignificant} argue that NGC 6121, the only LKH metal-rich GC known before NGC 6642, should have formed near the centre of the LKH potential well, and should therefore be the LKH cluster with the lowest orbital energy today, at odds with what observed. Although this can be interpreted as an inconsistency only by assuming that the orbital energy is perfectly conserved \citep[and, particularly in the inner galaxy, this is not the case, see e.g.][]{2023Pagnini,2024Chen,2025W}, NGC 6642 provides an even alternative solution to the problem, as it is the LKH cluster with the lowest Jacobi energy value. This finding thus reinforces the interpretation of a well defined AMR sequence for LKH, that is clearly separated both from the in-situ and from the GSE AMRs.  \\
Pal 1 is instead of a less immediate interpretation. Its age, when compared to that of the other in-situ GCs at similar metallicity, appears too young (by more than 2.5 Gyr) to share an in-situ origin. On the other hand, though, its high metallicity hardly matches that of accreted progenitors. The only one reaching such a high [M/H] would be Sagittarius. However, Sagittarius is a recent, ongoing event, whose debris has not yet phase-mixed entirely, so that its GCs are still located nearby its main body or its streams on the plane of the sky. Pal 1 is tens of degrees away, and therefore its association with Sagittarius, as well as with any other accreted progenitor, is highly unlikely. Just as a speculative interpretation, the only chance we see to explain such a young age and high metallicity is for Pal 1 to have formed in-situ, but from a gas that was diluted by a more metal-poor merger event. Since Pal 1 formed at the same time as NGC 6791, but with a much lower [M/H], our interpretation implies that such a dilution had to be a rather localized event. This is however consistent with the peculiar dynamical properties of Pal 1; it is the in-situ cluster with the largest apocentre (apo $\sim$ 18.37 kpc), similarly to Pal 10, E3, and ESO93-SC08. \\
Interestingly, \cite{Naidu_2020} discovered that Pal 1 shares many chemo-dynamical properties with Aleph, a previously unknown prograde substructure they identified in the integrals of motion space \citep[see also][]{De_Leo_2026}. \cite{Horta_2022_Al} later showed that this structure occupies the same region in the Mg-Fe plane as the metal-poor component of the low-$\alpha$ disc, which combined with its disk-like orbits suggests an in-situ origin. \\
This is in agreement with our interpretation, and to lend further support to it we use an additional chemical argument concerning [Eu/Si], that has been shown by \cite{2024Monty} to be a sensitive tracer of the origin of GCs. Specifically, \cite{Sakari_2011} yield an average ratio of [Eu/Si]$\sim$ 0.34 by means of high-resolution spectroscopy. Extending the relation shown by \cite{2024Monty} for the trend of [Eu/Si] vs [Fe/H] to [Fe/H]$\sim$ -0.6 observed for Pal 1, the value of [Eu/Si]=0.34 lies in between the in-situ and the accreted curves, consistent with the hypothesis of gas dilution.\\
Following the re-classification of NGC 6642 as a member of the LKH population, our final in-situ catalogue, which will be used in the next sections, includes a total of 33 clusters.
\subsection{The Age Distribution Function (ADF) and its relation with the MW merger history}\label{sec:adf}
\begin{figure}[ht!]
\label{fig:gmm}
\includegraphics[width=1\linewidth]{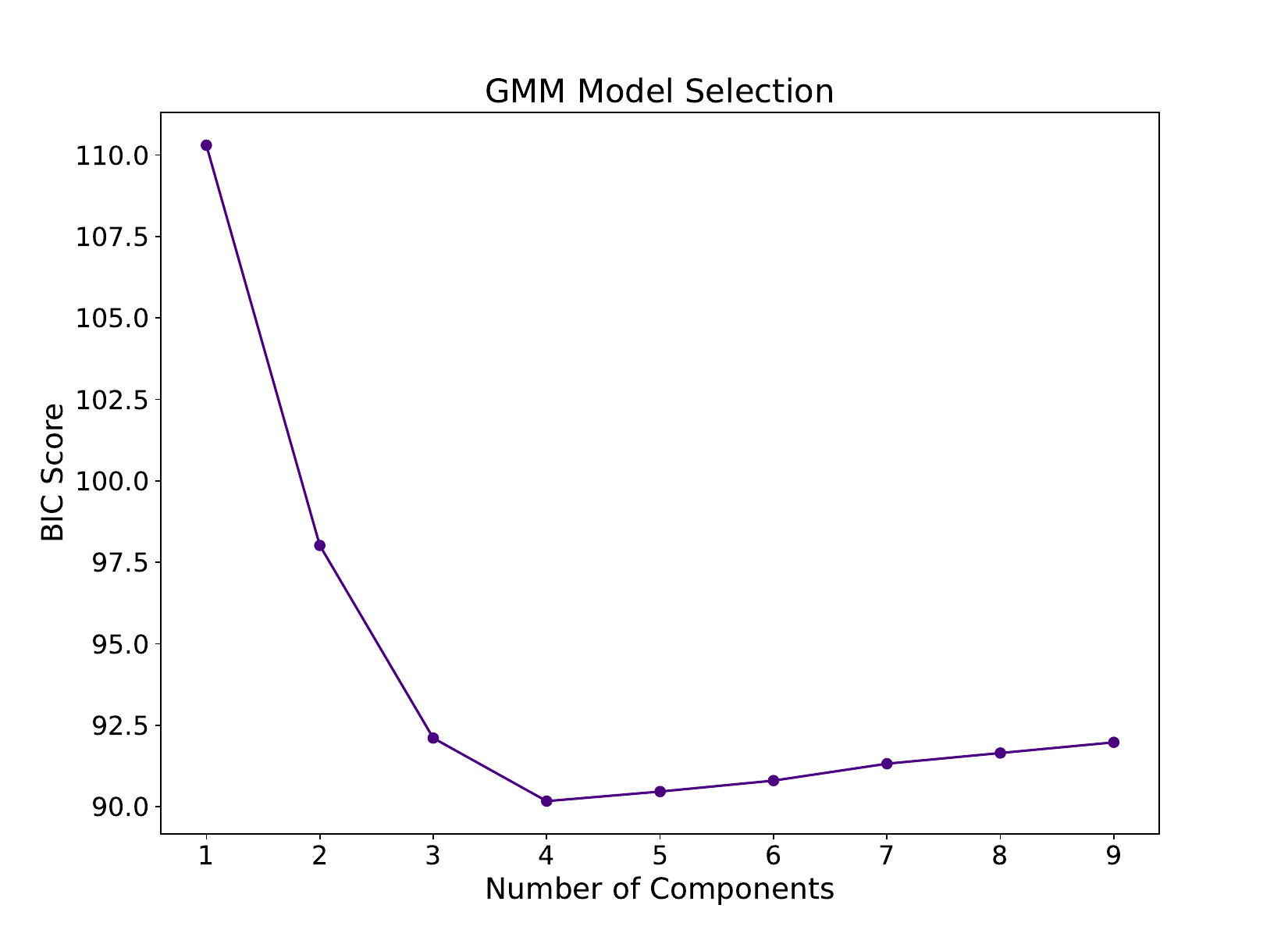} 
\caption{Bayesian Information Criterion (BIC) for the sample of random data points generated based on our real data set using a skewed Gaussian function that takes into account the asymmetric errors. As can be seen, the best BIC (minimum value) is obtained for a model with 4 Gaussian components.}
\label{fig:bic}
\end{figure}

As a further piece of information we can extract from the MW AMR is the timeline of star formation episodes in our Galaxy. To do so, in this section we investigate the multi-modality of the ADF, based on the derived GC ages.\\
To estimate statistically the most likely number of GC formation episodes building up the AMR, we employed a Gaussian Mixture Model (GMM) decomposition. We first generated, for each observed GC, a random realization of 50 data points, drawn from a skewed-normal distribution that incorporates the upper and lower uncertainties. We then applied a GMM to the combined synthetic sample: this procedure ensures that the underlying clustering of GCs towards specific distribution peaks is reflected more accurately than applying it directly to our observed age distribution without accounting for their associated errors. The GMM was evaluated for a varying number of components ranging from 1 to 10, and for each model we computed the Bayesian Information Criterion (BIC), which attains its minimum value when the probability is the highest. The resulting BIC distribution is shown in Fig. \ref{fig:bic}, where we can see that the most probable number of clusters corresponds to a model with four Gaussian components. \\

\begin{figure}[ht!]
\centering
\includegraphics[width=1\linewidth]{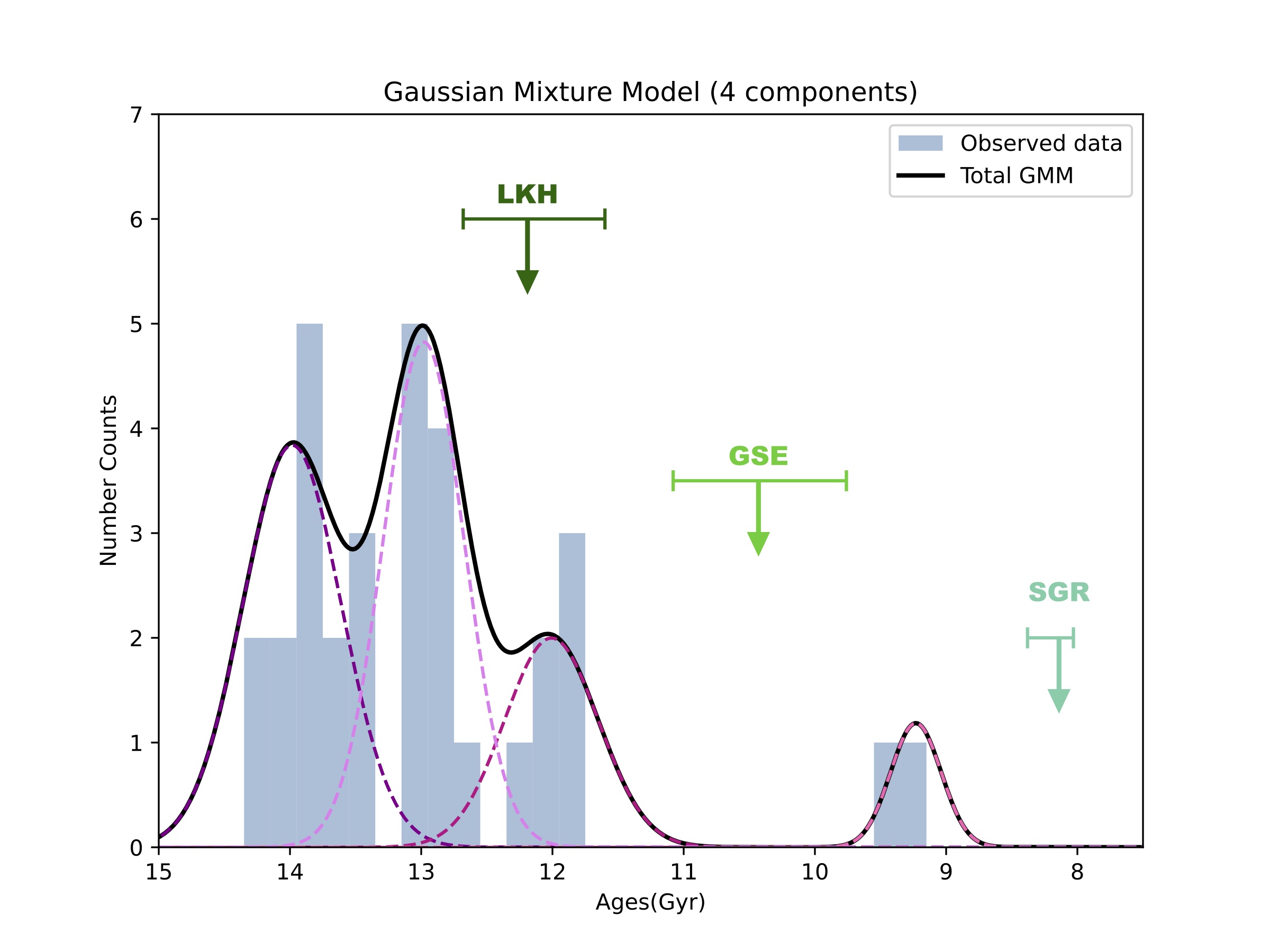}
\caption{Plot of the 4 Gaussian functions arising from the GMM compared to the histogram containing our real data set. We found four peaks at 13.97, 13.02, 12.00, and 9.25 Gyr, in order of decreasing look-back time. The green arrows with the associated errorbars show the timescales at which the MW merged with LKH, GSE, and the Sagittarius (Sgr) dwarf galaxy, respectively, as derived in the CARMA project (see M26 for LKH and GSE, and \cite{2026Aguado} for Sgr).}
\label{fig:gmm_hist}
\end{figure}
Figure \ref{fig:gmm_hist} shows the four Gaussian components over-plotted to the histogram of our observed data, which are grouped in bins of 0.2 Gyr. The peaks of GC formation episodes, that we interpret as corresponding to the mean age of the four components, are found, in order of decreasing look-back time, at 13.97, 13.02, 12.00, and 9.25 Gyr, respectively.\\
The plot also indicates (green arrows with errorbars) the accretion time for the most massive merger events as derived in the CARMA project, namely LKH (which has been estimated to have been accreted by the MW at $t=12.19^{+0.49}_{-0.59}$ Gyr, see M26), GSE (accreted at $t=10.43^{+0.65}_{-0.67}$ Gyr, see M26), and Sagittarius (accreted at $t=8.14^{+0.24}_{-0.11}$ Gyr according to \cite{2026Aguado}).
To interpret these results we relied on the findings of cosmological simulations (Di Cintio et al. 2020, Sotillo-Ramos et al. 2022, Orkney et al. 2022) which have consistently demonstrated that intense episodes of star/GC formation are triggered by an infalling, gas-rich merger, such as the ones we are discussing here. There is still ambiguity in the literature about the exact moment when the trigger takes place. In general, it is safe to state that in-situ star formation peaks at around the pericentric passages of the incoming merger, and can be enhanced during the whole interaction (e.g., Ruiz-Lara et al. 2020, Valenzuela et al. 2024). 
According to this, we tentatively associated each peak of in-situ GC formation to the closest successive merger event. In this sense, it is interesting to notice that LKH appeared to have merged with the MW at an epoch that is very close to the second peak of GC formation at 13.02 Gyr, within the uncertainties. Instead, if we look at later merger events like GSE and Sagittarius, the observed data suggest a larger delay between the peaks of in-situ GC formation episodes and the accretion time of these mergers. This delay changes from $\Delta t$ (MW-LKH)$=0.8^{+0.5}_{-0.6}$ Gyr, to $\Delta t$ (MW-GSE)$=1.6\pm 0.7$ Gyr and to $\Delta t$ (MW-Sag)$=1.2^{+0.2}_{-0.1}$ Gyr.\\
We tested the robustness of the Gaussian decomposition and of the location of each peak by means of a bootstrap test with replacement (see Appendix \ref{app:bootstrap} for details), finding high statistical significance especially for the three youngest peaks.\\

\begin{figure}[t]
    \centering
    \includegraphics[width=1\linewidth]{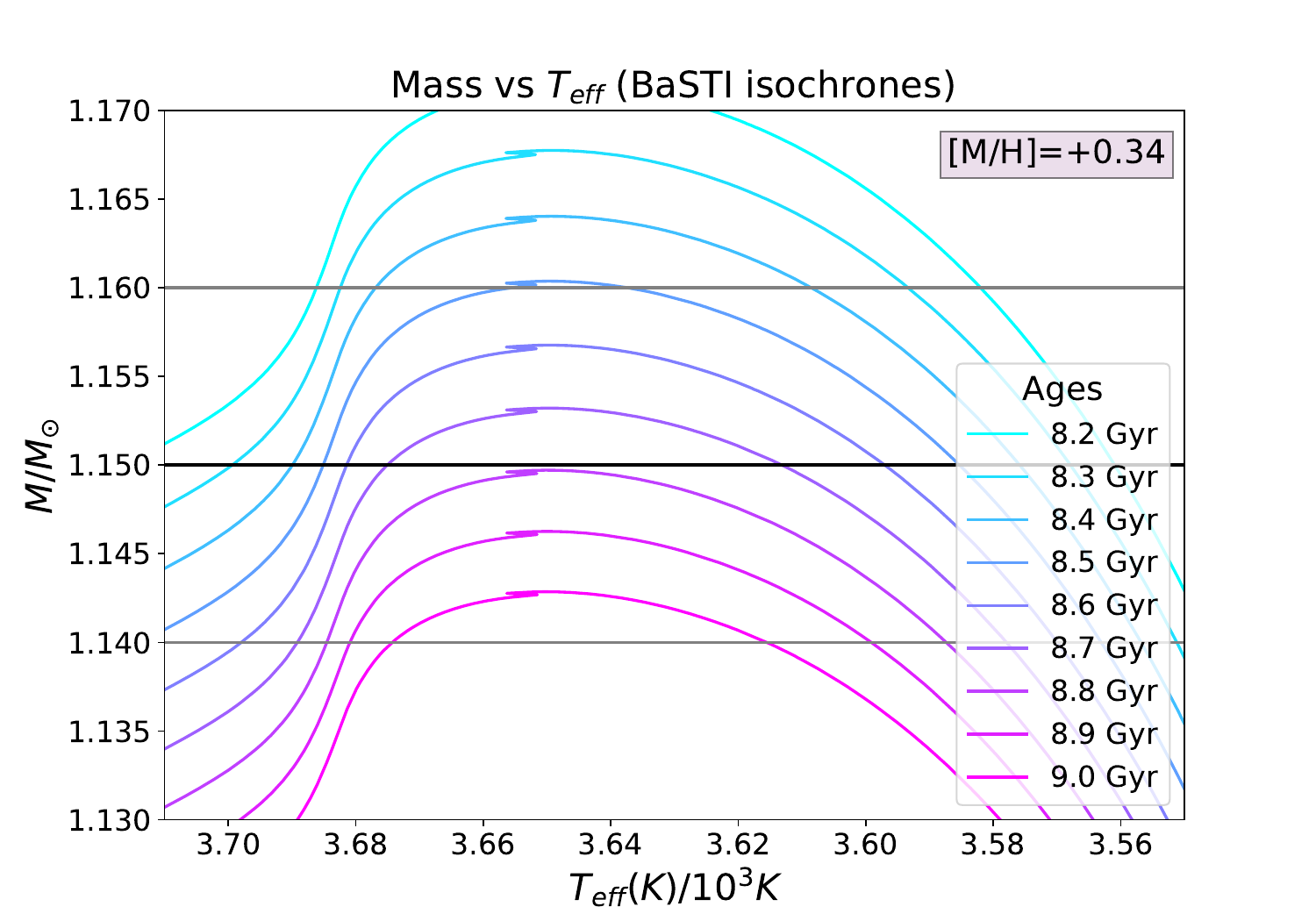}
    \caption{The plot of mass versus effective temperature shows the intercept of the asteroseismic mass of $1.15 \pm 0.01M_{\odot}$ for RGB stars in NGC 6791 with a set of BaSTI isochrones of different ages along the RGB at the fixed metallicity of [M/H]=+0.34 estimated for this GC. The mass is shown as a black solid line, along with its uncertainties (gray solid lines).}
    \label{fig:massvste}
\end{figure}
Even though the associations between the GC formation peaks and the merger events require further confirmation, we make an attempt at validating our interpretation using the predictions from the E-MOSAIC cosmological simulations described in \cite{Hughes_2018}, H18 hereafter.
H18 defined the time of infall as the last time a satellite galaxy crossed the virial radius of the main galaxy, and as time of the merger the moment in which the satellite is effectively cannibalized. According to their Fig. 9, the later the merger is, the larger the delay between these two timescales is. In other words, earlier mergers, that took place when the accreted mass was similar to that of the host, are almost instantaneous. Their gas is immediately stripped and star formation is halted. This scenario is, within the errors, compatible with what we observe for the LKH merger event.\\
Instead, H18 showed that later mergers were able to retain enough gas and pressure to continue to form GCs even after their first interaction with the main galaxy \citep[see, e.g.,][]{Gonz_lez_Koda_2025,Aguado_Agelet_2025} so that the time it takes for them to completely merge after infall is larger. This is exactly what we observe, within the uncertainties associated with the accretion time of the mergers.\\
We point out that the definition of "infall time" is not univocal in the literature, as it is inherently difficult to uniquely define what constitutes the first interaction between a satellite and its host. Depending on the involved physical processes, this baseline can range from the crossing of the host's virial radius to the onset of tidal stripping. In light of all of these considerations, we propose a scenario where the peaks of in-situ GC formation were triggered by the first pericentric passage of each accretion event, prior to the moment at which GC formation in the merger itself is halted after gas stripping (corresponding to the parameter t$_f$ in the AMR models of the CARMA framework, see Sect. \ref{sec:model}), which takes place with larger and larger delay the more the evolution of the Galaxy goes on.
\subsection{A "seismic-grounded" absolute age scale} \label{sec:absolute}
Although isochrone fitting can be highly precise when determining relative ages, high accuracy on the absolute age scale is more challenging to achieve. In fact, this strongly depends on poorly known aspects of the stellar evolution that are used as ingredients in the models, as well as on the uncertainties in the  bolometric corrections, and in the properties of the stellar population \citep[e.g.][]{2025Ying}. So far, the CARMA project has always interpreted ages in a relative sense, but given the cosmological relevance of in-situ GCs, in this work we make an attempt to calibrate our absolute age scale to an empirically calibrated time reference that partly overcomes the problems associated with ages derived from the fitting of isochrones. \\
Asteroseismology is a compelling tool for calibrating such an age scale, as it provides an independent measure of the age in star clusters. In fact, the observed oscillation modes of red giant stars allow their mass to be determined with high precision and, consequently, their age \citep[see,e.g.][]{2010aste.book.....A,Chaplin_2013}. As pointed out in \cite{Lebreton_2014}, even though still model dependent, the asteroseismic age scale is fairly more accurate than that obtained with traditional methods like isochrone fitting (see Sect. \ref{sec:caveats} for the caveats). \\
In this section we adjust the zero-point of the absolute age scale for GC formation to the asteroseismic absolute age scale by comparing the isochrone fitting results with asteroseismic studies using NGC 6791 as reference cluster.
The choice of NGC 6791 is dictated by the fact that it is the only star cluster\footnote{NGC 6791 is unanimously referred to as a massive, old open cluster, given the absence of multiple populations \citep{Bragaglia_2014,Dondoglio_2022}.} in our sample for which a sufficiently precise asteroseismic measurement of the mass of its stars exists. Moreover, it hosts at least three eclipsing binaries (EB). The importance of some of these EB comes from they location in the CMD (near the turn-off MS and in the RGB phase), providing independent and more accurate mass estimates than other methods (high accuracy is indeed crucial to achieve a better age scale). For these reasons, several studies have been carried out with the aim of estimating its age in an accurate way \citep[e.g.][]{Brogaard_2012}.\\
In order to determine the seismic-grounded absolute zero-point of our age scale, we proceeded as follows.
First of all, we looked for an asteroseismic estimate of the mass for stars along the RGB of NGC 6791. By analysing the Kepler light curves of 27 RGB stars, \cite{McKeever_2019} found an average mass of $1.15\pm 0.01M_{\odot}.$ This value is in excellent agreement with the mass on the low RGB estimated by fitting the location of three eclipsing binaries in the mass-radius space by \cite{Brogaard_2011}. The tight consistency of these mass estimates suggests that, for NGC 6791, mass loss is not the main "player" in the lower part of the RGB phase, as expected from models. \\
Other studies \citep[e.g.][]{2011Basu,2012Miglio} reported slightly different values for the asteroseismic mass of RGB stars in NGC 6791, and this difference depends also on the adopted surface correction terms \citep[][]{J_rgensen_2020}.
However, the surface correction terms applied by \cite{McKeever_2019} are sufficiently flexible to account for both incorrect modelling of the near-surface layers, and to result in less scatter in mass values around the mean.\\
At this point, we needed to find which among our used BaSTI models at the metallicity we found for NGC 6791 of [M/H]$\simeq$+0.34 has a mass at the low RGB of M=1.15 $M_{\odot}.$ The results are shown in Fig. \ref{fig:massvste}, where we plot a grid of BaSTI isochrones at fixed [M/H]=+0.34 and with ages varying with a step of 100 Myr in the mass vs effective temperature space. \\
As it can be appreciated in the plot, the isochrone intercepting the mass of $1.15 M_{\odot}$ at the base of the RGB corresponds to that having an age of 8.5 Gyr. Accounting for the upper and lower errors on the mass estimate, we estimated an absolute age for NGC 6791 $t=8.5\pm 0.3$ Gyr, and therefore we changed the zero point of all our GC ages by an amount $t_{iso}-t=0.65$ Gyr, where $t_{iso}=9.15$ Gyr is the age derived from the isochrone fitting of NGC 6791. With this new prescription, the oldest GC in the total MW in-situ sample is NGC~4372, with an age of $13.83\pm 0.12$ Gyr.\\
Even though NGC 6791 is the only cluster for which precise asteroseismic masses have been derived from individual frequencies, its regime of high metallicity and young age with respect to ordinary GCs may cast doubts on the reliability of our absolute age calibration. For this reason, we validated our analysis in a metallicity and age range more suitable to "classical" GCs. Therefore, we repeated the analysis on the GC NGC~6121 - albeit its asteroseismic masses have been derived by means of scaling relations - for which a CARMA age have been derived in M26. Its analysis resulted in an offset of $t_{iso}-t=0.54\pm0.5$ Gyr that is remarkably consistent with the one obtained from NGC~6791 (see Appendix \ref{app:m4} for details), thus supporting the reliability of our absolute age calibration also in a regime of age and metallicity more typical of standard GCs.
\begin{figure}[ht!]
\includegraphics[width=1\linewidth]{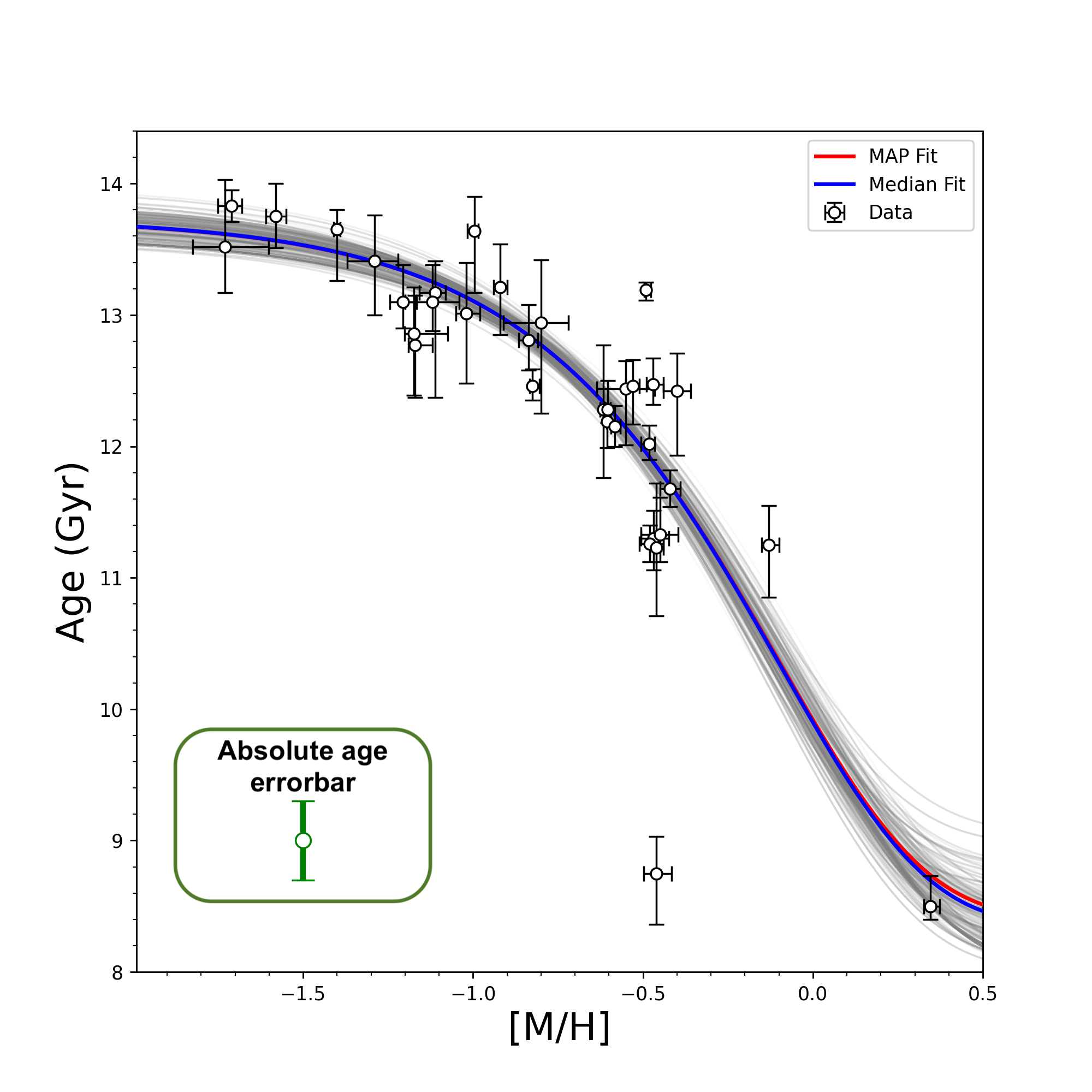} 
\caption{Best-fit function to our data (white circles) for a constant SFR model. The median fit, corresponding to the 50\textsuperscript{th} percentiles of the posterior distribution (solid blue line) and the MAP fit (solid red line) are shown. In the lower left corner, we report the errorbar associated with the absolute age re-calibration.}
\label{fig:fit_exp}
\end{figure}
\subsection{The age of the Universe} \label{sec:h0}
\label{sec:ageuni}
After having re-calibrated all the ages of the in-situ GCs according to the offset derived in Sect. \ref{sec:absolute}, we fitted the new AMR in a Bayesian framework by means of a MCMC approach. We used both the AMR models illustrated in Sect. \ref{sec:model}, and we adopted a logarithmic prior $\log P$ that consisted of three terms:
\[
\log P=\log P_{M}+\log P_{t}+\log P_{\eta},
\]
where, $P_{M}$ is associated with the MW stellar mass, and is uniformly distributed in the range $10^{9}< M_{*}<10^{13}M_{\odot};$ $P_{t}$ is the prior on $t_{i}$ and $t_{f},$ and is also flat in the interval between $8<t_{i},t_{f} <15$ Gyr; $P_{\eta}$ is related to the asymmetric intrinsic dispersion on the AMR, and has a Gaussian shape with $\sigma=100$ Myr (which is the typical dispersion measured around the observed data at fixed metallicity).\\
The total logarithmic likelihood function $\log L$ is represented as the sum of two terms:
\[
\log L=\log L_{t}+ \log L_{z},
\] 
where the first one is associated with the fit along the age direction, and the latter with the fit along the metallicity direction. 
Our fitting procedure took into account the asymmetric uncertainties on the data.\\
We performed the fit using both the AMR models described in Sect.~\ref{sec:model}. To estimate which of the two represented the best fit for our data, we computed the BIC in both the MCMC runs, and found BIC(declining SFR)-BIC(constant SFR)$>20$, which according to the Kass \& Raftery criterion \citep[][]{Kass01061995} strongly favours the model with constant SFR. Therefore, we based our subsequent analysis on the results provided by this model, which gave an initial timescale for GC formation of $t_{i}=13.59^{+0.13}_{-0.12}$ Gyr. We remark nonetheless that the fit of the model with declining SFR provided extremely similar results, with t$_i$ varying by less than the 2\%. The resulting best fit is shown in Fig. \ref{fig:fit_exp}. The solid blue line refers to the median values, taken as the 50\textsuperscript{th} percentiles of the posterior distribution, while the red line represents the Maximum a Posteriori (MAP) estimate of the Bayesian analysis. \\
We now focus on the key question we want to answer in this section, which is whether we can estimate the age of the Universe based on our observed GC AMR. By adding in quadrature the uncertainty of $\pm 0.3$ Gyr coming from the absolute age re-calibration to our value of $t_i,$ we obtained an initial timescale for GC formation of $t_{GC,i}=13.59^{+0.33}_{-0.32}$ Gyr.\\
However, this moment does not coincide with the age of the Universe, as GC formation is not instantaneous, but should occur after a delay $\Delta t.$ Such a delay has been estimated by \cite{Jimenez_2019}. In this work, the authors used the ages of the oldest stars to infer a $\Delta t$ value of $0.2 \pm 0.1$ Gyr. By applying this offset, and accounting for this additional source of error, we obtain for the age of the Universe a value of $t_{U}=13.79\pm 0.34$ Gyr.
\subsection{The Hubble constant}
\begin{figure}[]
\centering
\includegraphics[width=1\linewidth]{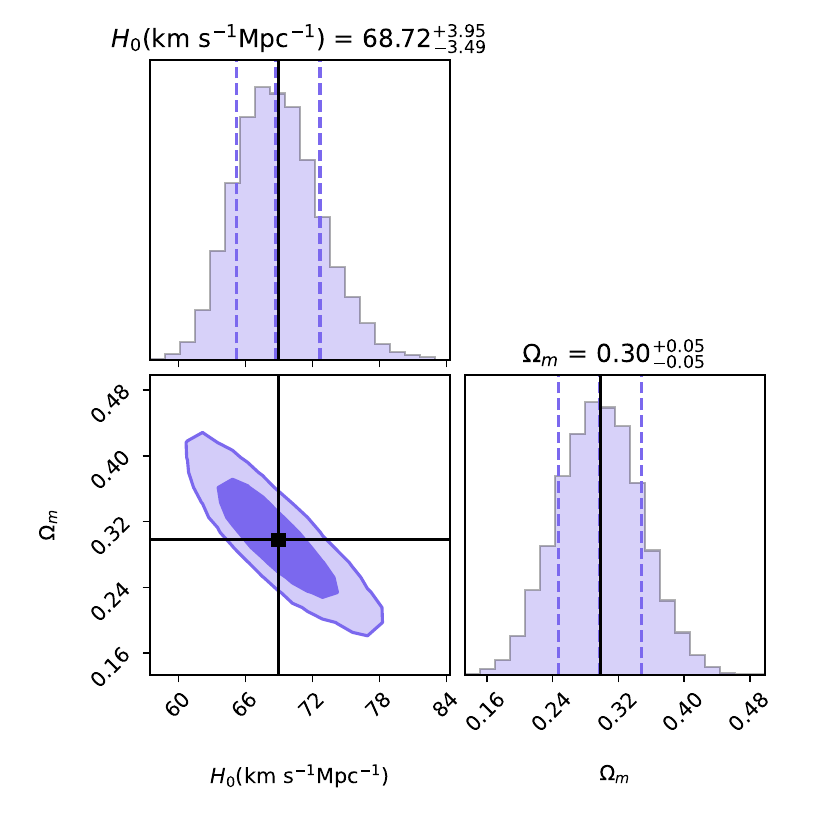} 
\caption{Corner plot showing the posterior distribution of the MCMC procedure used to find the numerical solution of Equation \ref{eq:h0}. The solid lines represent the median values, while the dotted lines display the 16\textsuperscript{th} and the 84\textsuperscript{th} percentiles of the distribution.}
\label{fig:h0}
\end{figure}

\begin{figure}[h!]
\centering
\includegraphics[width=1\linewidth]{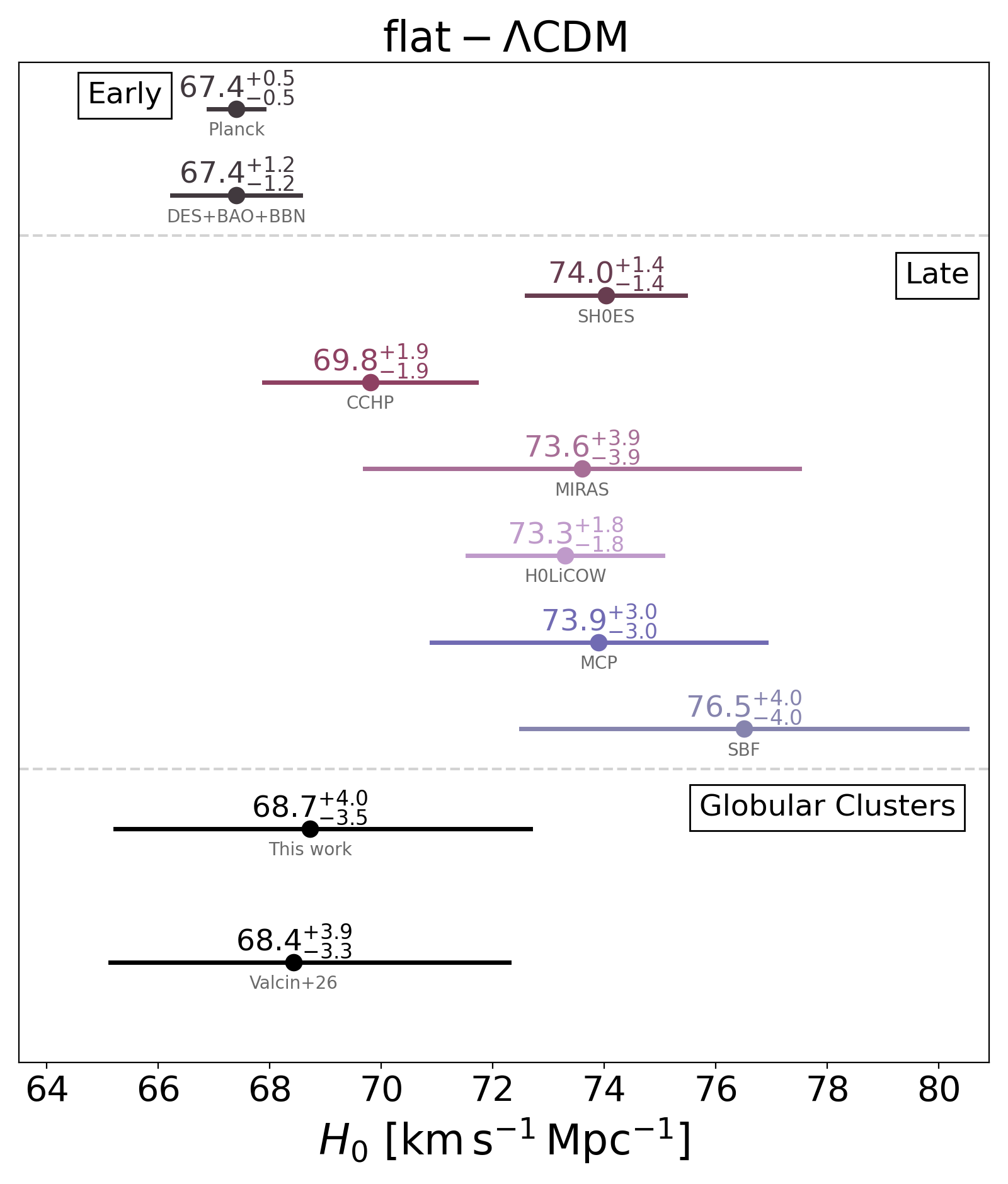}
\caption[]{The plot compares the Hubble constant derived from GCs with estimates from different probes. Upper Panel. Results from early-universe observations: Planck collaboration \citep{Planck_2020}; Dark
Energy Survey (DES) + BAO + Big Bang Nucleosynthesis (BBN) \citep{Beutler_2011,Ross_2015,Alam_2017,Troxel_2018,Abbott_2018,2019Macaulay}. Middle Panel. Late-universe observations using different cosmological probes, in order: a late universe combination of Supernovae-$H_0$ for the Equation of State
of Dark energy \citep[SH0ES][]{2016SH0ES}; Status Report on the Chicago-Carnegie Hubble Program \citep[CCHP][]{freedman2025statusreportchicagocarnegiehubble}; Mira variables \citep[MIRAS][]{Huang_2020}; H0 Lenses in COSmological MOnitoring of GRAvItational Lenses (COSMOGRAIL) Wellspring \citep[H0LiCOW, see][]{2019h0licow1,h0licow}; the Maser Cosmology Project MCP \citep{Pesce_2020}; the Cepheids surface
brightness fluctuations method \citep[SBF][]{2019Verde}. Lower Panel. GC-based values of $H_0$, from the age of the Universe derived in this work and in \cite{valcin2026ageuniverseglobularclusters}, respectively.\footnotemark}
\label{fig:tension}
\label{fig:h0_tension}
\end{figure}

In a generic cosmological model, the Hubble constant $H_0$ can be computed as:
\begin{equation} \label{eq:h0gen}
    H_{0}=\frac{A}{t} \int_{0}^{z_{f}} \frac{1}{(1+z')E(z')} dz', 
\end{equation}
where $E(z)=H(z)/H_{0}$, $t$ is the age of an object formed at a redshift $z_{f},$ and A=977.78 is the factor needed to convert from Gyr (units of $t$) to km s\textsuperscript{-1} Mpc\textsuperscript{-1} (units of $H_0$). In the limit $z \rightarrow \infty,$ $t$ converges to the age of the Universe $t_{U}.$\\
In a flat $\Lambda$CDM cosmological model, Eq. \ref{eq:h0gen} reduces to: 
\begin{equation} \label{eq:h0}
    H_{0}=\frac{A}{t} \int_{0}^{z_{f}} \frac{1}{(1+z')} [\Omega_{M}(1+z')^{3}+\Omega_{\Lambda}]^{-1/2}dz',
\end{equation}
where $\Omega_{M}$ and $\Omega_{\Lambda}=1-\Omega_{M}$ are the density parameters associated to the total matter and dark energy, respectively.
Putting together several low redshift determinations of $\Omega_{M},$ \cite{Jimenez_2019} found a joint prior of $\Omega_{M}=0.3 \pm 0.02$. This latter, combined with our result for $t_{U},$ can be used to find a numerical solution to Eq. \ref{eq:h0}. This was done by means of a MCMC algorithm, and adopting a Gaussian log-likelihood function for $t_{U}.$ We imposed conservative priors both on $H_{0},$ which we left free to vary between 40 and 100 $\text{km s}^{-1} \text{Mpc}^{-1},$ and on $\Omega_{M},$ for which we adopted a Gaussian prior of 0.3 with $\sigma=0.05.$ The final solution converged to $H_{0}=68.72^{+3.95}_{-3.49} \; \text{(stat.)} \; \text{km s}^{-1} \text{Mpc}^{-1},$ and it is shown in Fig. \ref{fig:h0}, where we plot the posterior distributions for $H_{0}$ and $\Omega_{M}$ resulting from the MCMC procedure.\\
    We have also estimated the impact of possible zero-point systematics in the age scale set by the stellar models. To this end, we took advantage of the analysis by \cite{Ying_2025}, who evaluated the effect of realistic uncertainties on many input parameters/physics (see their Table 1) on the absolute ages of stellar models in the GC's regime. Based on their results, we considered a conservative systematic zero-point uncertainty of $\pm 0.5$ Gyr on our ages, which translates to a systematic error on $H_{0}$ of $\sigma _{\text{sys}}=^{+2.65}_{-2.32} \; \text{km s}^{-1} \text{Mpc}^{-1}.$\\
To compare our results with the recent findings from \cite{valcin2026ageuniverseglobularclusters}, who also estimated the age of the Universe using the absolute age of GCs, we computed $H_0$ using the same methodology described here, but including their estimate as input. The corresponding value we found is $H_{0}=67.93^{+3.86}_{-3.29} \; \text{(stat.)} \; \text{km s}^{-1} \text{Mpc}^{-1},$ in excellent agreement with our results.\\
This comparison is particularly relevant in the current debate about the so-called Hubble tension \citep[see][for a recent review]{2024_Verde}.  It is now established that while early cosmological probes enabled inferring $H_0$ with typical values of $H_0\approx67 \; \text{km s}^{-1} \text{Mpc}^{-1}$, local tracers systematically provide larger values (around 73 km s\textsuperscript{-1}Mpc\textsuperscript{-1}), and they differ at a value greater than 5 sigma (as also shown in Fig. \ref{fig:h0_tension}). If the tension is confirmed, and not due to possible systematics, it would require either a complete revision of the $\Lambda$CDM model or the discovery of a new physics at the early times. In this context, independent cosmological probes are important to better understand the expansion history of the Universe, and GCs can provide an interesting class to directly test the “Hubble tension” \citep[see also][]{cimatti2023revisitingoldeststarscosmological}).\\
We derived the age of the Universe from GCs independently from \cite{valcin2026ageuniverseglobularclusters}, but the two values agree remarkably well (see Sect. \ref{sec:absolute}). As a result, the median value of $H_{0}$ that these $t_{U}$ estimates lead to closely match the $H_{0}$ values from the early Universe probes, even though within the uncertainties our results are still consistent with $H_{0}$ as derived from local probes. 
Waiting for future dedicated effort (e.g., from additional asteroseismic data) to shrink the uncertainties, such a result opens a question regarding the nature of GCs as tracers of $H_{0}$: should they be considered as local probes or not? Being the oldest objects in the resolved Local Universe, the answer is not straightforward and the discussion is still open \citep[see e.g.][]{2022Vagnozzi}. If GCs were considered a genuine early Universe probe, our estimate is in agreement with the results from the CMB and BAOs (Fig. \ref{fig:h0_tension}), within the assumed $\Lambda$CDM model. It should be noted, however, that while our median value aligns more closely with early universe measurements, it remains compatible with late universe probes within about 1.1$\sigma.$ Our considerations should therefore be taken with caution. As suggested by \cite{2024_Verde} (Review), cosmic chronometers (such as GCs) could be competitive probes in shedding light on the “Hubble tension”, but they still need to be scrutinized for being an asset in this cosmological application.
\section{Discussion and conclusions} \label{sec:conc}
In this work, we analysed the AMR of 34 in-situ star clusters. We complemented previous CARMA age estimates with new isochrone fitting results for 18 GCs, finding that among the entire sample only 2 GCs lie in a region of the AMR plane that seems inconsistent with an in-situ origin, namely NGC6642 and Pal1. \\
While NGC6642 is re-classified as belonging to the LKH event (M26), and thus becomes its lowest energy, metal-rich member, we propose Pal 1 to have still formed in-situ, but from a metal-poorer gas cloud diluted by the interaction with a metal-poorer merger-like event. In this sense, Pal 1 can be considered as related to the Aleph \citep{Naidu_2020} stellar sub-structure \citep[see also][]{De_Leo_2026}.\\
We studied the age distribution function of our GC sample using a Gaussian Mixture Model that suggested the presence of 4 Gaussian distributions corresponding to four separate episodes of GC formation, at 13.97, 13.02, 12.00, and 9.25 Gyr. While the first and oldest peak is expected to follow naturally from the evolution of the Galaxy, we explored the possibility for the three later peaks to be linked to the infall of the most significant merger events currently known.\\
We find that each of the found peaks of in-situ GC formation can possibly be related to the accretion of LKH, GSE and Sagittarius, respectively. In particular, while the delay between the second peak at 13.02 Gyr and the LKH merger event appears small (<1 Gyr), the delay between the subsequent peaks and more recent mergers like GSE and Sagittarius is larger, exceeding 1 Gyr. Such a trend matches the predictions from the E-MOSAICS cosmological simulations in H18, which find that while early mergers extinguish their gas almost immediately, later mergers can keep gas even after their first interaction with the MW (which triggers a peak in the in-situ GC formation) and can thus form GCs later on.
At this point, to calibrate our ages to a more accurate absolute scale, we used the asteroseismic mass of lower RGB stars in NGC 6791 from \cite{McKeever_2019}, who found $M=1.15 \pm 0.01 \; M_{\odot}.$ This mass value intercepts our BaSTI models at the metallicity determined by our isochrone fit ([M/H]=+0.34) at an age of 8.5 Gyr. This provides a 0.65 Gyr offset to re-calibrate our age scale, which was subsequently applied to the entire GC sample. \\
Finally, once the zero point of the age scale was adjusted, we fitted the new age-metallicity relation by means of a Markov-Chain-Monte-Carlo approach, and found an initial timescale for GC formation in the MW $t_{GC,i}=13.59^{+0.33}_{-0.32}$ Gyr. Following the prescription in \cite{Jimenez_2019} for the delay between the Big Bang and the formation of the first stars, we inferred a GC-based age of the universe  $t_{U}=13.79\pm 0.34$ (stat.) $\pm 0.5 $ (sys.) Gyr, corresponding to a Hubble constant of $H_{0}=68.72^{+3.95}_{-3.49}\; \text{(stat.)} ^{+2.65}_{-2.32}\; \text{(sys.)} \; \text{km s}^{-1} \text{Mpc}^{-1}.$ Such a value is in agreement with those provided by probes in the early Universe, though still compatible with late Universe measurements within $\sim1\sigma$, and poses the question on whether GCs should be considered as such, or as local tracers of $H_{0}.$
\footnotetext{We acknowledge Vivien Bonvin and Martin Millon for the plot available at \url{https://github.com/vbonvin/H0_tension}}
\subsection{Caveats about our asteroseismic absolute scale}\label{sec:caveats}
The most precise asteroseismic masses, like those used in this work \citep{McKeever_2019}, are obtained by fitting individual oscillation frequencies with stellar evolutionary and pulsation models \citep[see, e.g.][]{Lebreton2014,Rendle_2019,J_rgensen_2020,Li2020,Montalban2021,Jiang2021,Aguirre2022,Lundkvist2025,Reyes2025,Kjeldsen2025,Li2026,Kallinger2026}, which depend on the ability of stellar evolutionary and pulsation models to reproduce the observed oscillation properties and internal stellar structure.\\
Hence, our asteroseismic absolute scale is not entirely free from systematic errors due to the many assumptions in the evolutionary models. For example, these might include radiative opacities, mixing prescriptions, rotation, magnetic fields, mass loss, and the presence of frequency shifts due to poor modelling of near-surface layers \citep[known as surface effects, see e.g.][]{Christensen-Dalsgaard1988,Kjeldsen2008,Ball_2014,Li2023,Kallinger2026}.\\
However, seismic masses can be measured independently for each star, allowing for internal consistency checks against some of these biases. In fact, in star clusters, asteroseismic masses can be measured for stars with the same age and chemical composition (at least at first order) across different evolutionary phases. Outliers and discrepancies can thus be identified more robustly, helping to constrain the impact of binary evolution \citep[see, e.g.][]{Handberg2017,Brogaard_2021,2024Matteuzzi,Leitinger2026,Howell2026}, mass loss \citep[][]{Howell2022,2022Tailo,Howell2024,Brogaard_2024,Howell2025,Stello2026,Howell2026}, surface effects \citep[][]{J_rgensen_2020}, and limitations in the adopted stellar physics. \\
An additional caveat we want to highlight is the fact that the absolute age scale used in this paper relies on the asteroseismic masses obtained in the open cluster NGC 6791. This cluster is substantially more metal-rich than typical Galactic globular clusters
\citep[see e.g.][]{Peterson1998,Gratton2006,Origlia2006,Bragaglia_2014}. Therefore, we assume throughout the paper that the systematic offset between the isochrone-fitting-based and the seismic age estimates for NGC 6791 can be approximately extended to more metal-poor and older GCs. This is a working assumption driven by the fact that highly precise asteroseismic mass measurements based on individual oscillation frequencies only exist for this cluster, but should be verified when more asteroseismic data in GCs will be collected.\\
Despite this caveat, our paper provides one of the first attempts to quantitatively evaluate the impact of systematic uncertainties on age-dating using independent age tracers, and effectively provides a practical framework for evaluating systematic uncertainties in GC age estimates. Future space missions such as PLATO \citep{2025PLATO} or the HAYDN project \citep{2024haydn} will allow significant steps forward towards much more accurate and precise GC absolute ages.
\begin{acknowledgements}
We kindly thank the referee for the very constructive feedback, which helped us to improve the quality of the manuscript.
DM, SC, and EP acknowledge financial support from
PRIN-MIUR-22: “CHRONOS: adjusting the clock(s) to unveil the CHRONO-chemo-dynamical Structure of the Galaxy” (PI: S. Cassisi).
FA acknowledges
support by the Galician Regional Government under project  ED431C 2026/51 “GRC" and atlanttic: supported by the grant PID2023-147500OB-I00, funded by MICIU/AEI/10.13039/501100011033 and by ERDF/EU. CF acknowledges
support by the PRIN INAF 2023 grant ObFu N\textsuperscript{2}RED (PI: C. Fanelli). CG and SC acknowledge support from the Agencia Estatal de Investigación del Ministerio de Ciencia e Innovación (AEI-MCINN) under grant “At the forefront of Galactic Archaeology: evolution of the luminous and dark matter components of the Milky Way and Local Group dwarf galaxies in the Gaia era” with reference PID2020-118778GB-I00/10.13039/501100011033 and PID2023-
150319NB-C21/C22/10.13039/501100011033. CG also acknowledge support
from the Severo Ochoa program through CEX2019-000920-S. M.M. acknowledges support from the Agencia Estatal de Investigación del Ministerio de
Ciencia e Innovación (MCIN/AEI) under the grant “RR Lyrae stars, a lighthouse to distant galaxies and early galaxy evolution” and the European
Regional Development Fun (ERDF) with reference PID2021-127042OB-I00.
CS and DM acknowledges financial support by the Italian Ministry of University and Research (grant FIS2023$-$01611, CUP C53C25000300001).
Co-funded by the European Union (ERC-2022-AdG,
”StarDance: the non-canonical evolution of stars in clusters”, Grant Agree-
ment 101093572, PI: E. Pancino). Views and opinions expressed are however those of the author(s) only and do not necessarily reflect those of the Euro-
pean Union or the European Research Council. Neither the European Union nor the granting authority can be held responsible for them.
Based on observations with the NASA/ESA HST, obtained at the Space Telescope Science Institute, which is operated by AURA, Inc., under NASA
contract NAS 5-26555. MM acknowledges support from the ERC Consolidator Grant funding scheme (project ASTEROCHRONOMETRY, \url{https://www.asterochronometry.eu}, G.A. n. 772293).
\end{acknowledgements}

\bibliographystyle{aa}
\makeatletter
\makeatother

\bibliography{ref}
\begin{appendix}\onecolumn
\section{Comparison with the literature} \label{app:comparisons}
\begin{figure}[th!]
\begin{subfigure}{0.5\textwidth}
\includegraphics[width=1\textwidth]{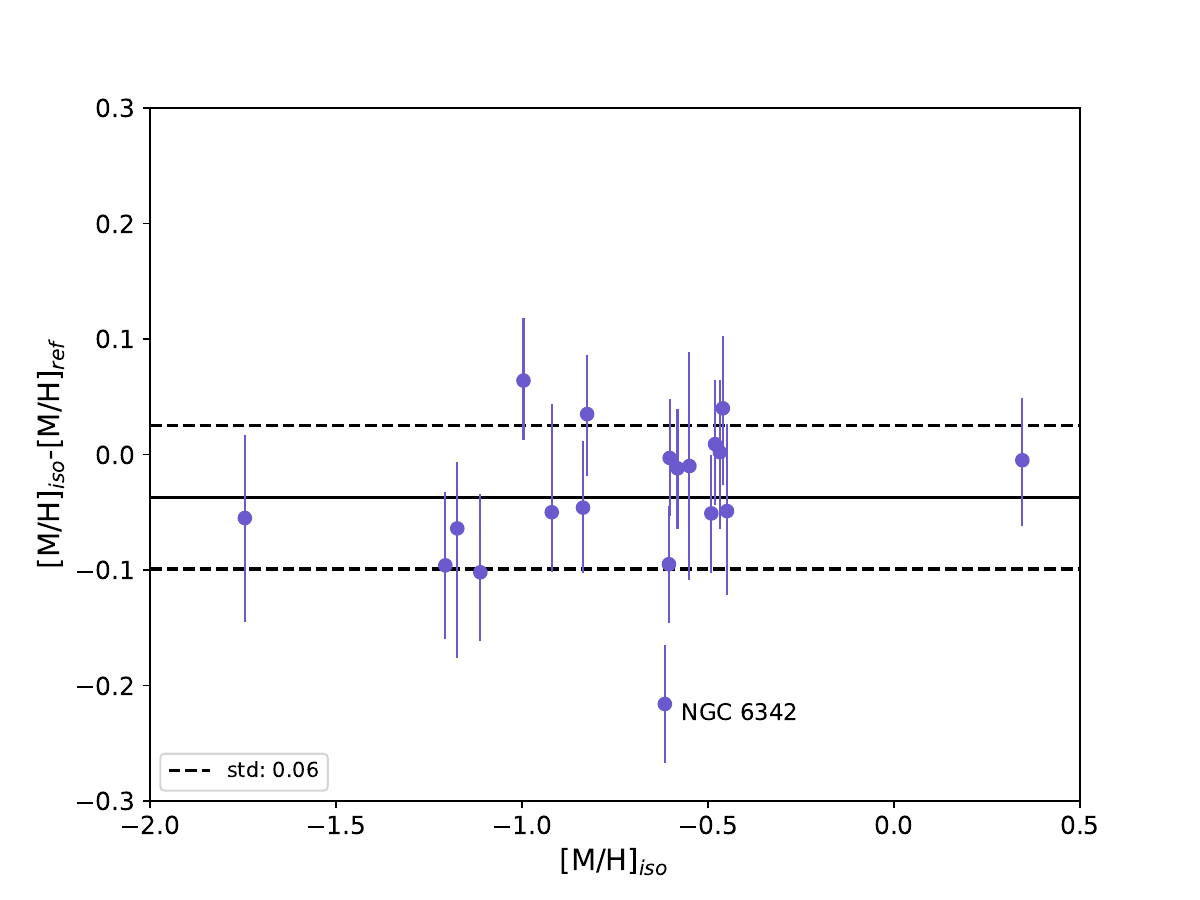} 
\caption{}
\end{subfigure}
\begin{subfigure}{0.5\textwidth}
\includegraphics[width=1\textwidth]{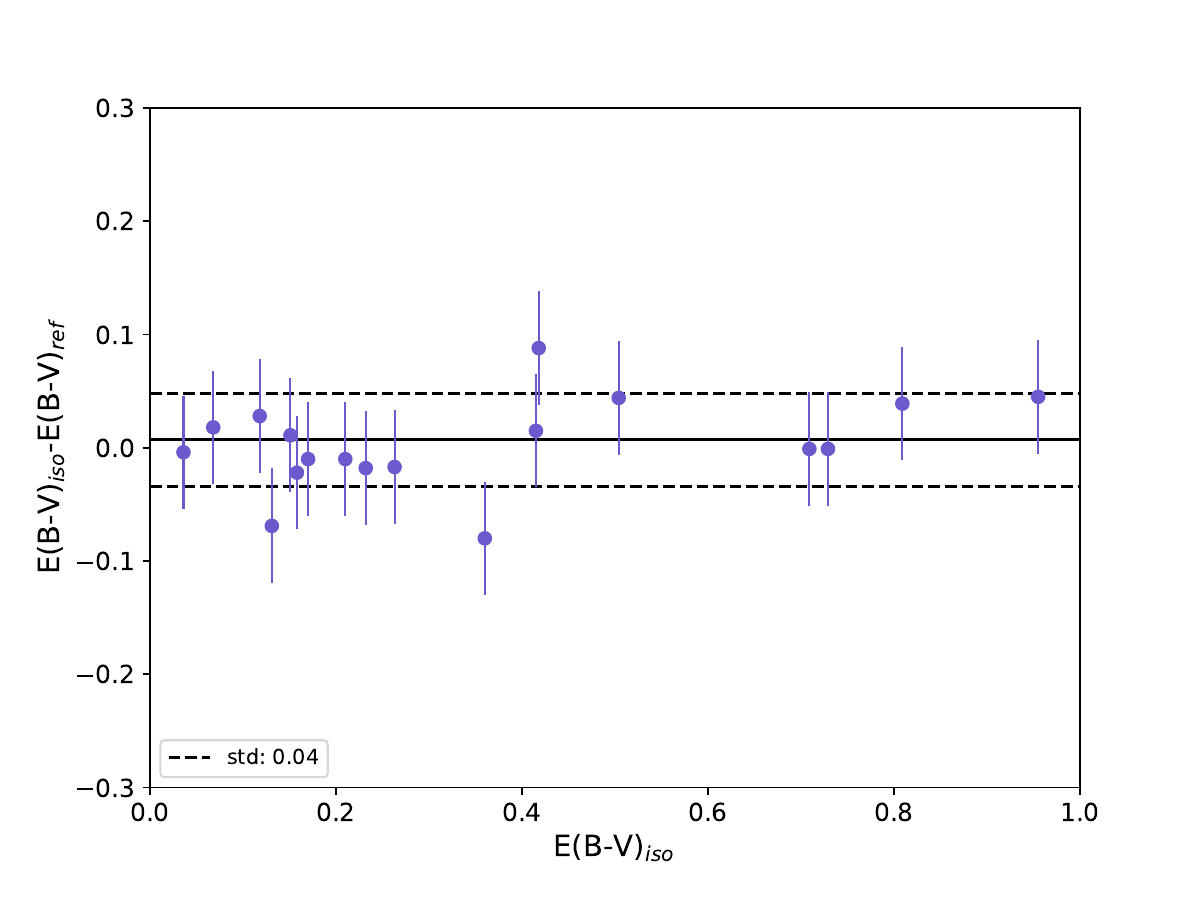}
\caption{}
\end{subfigure}
\begin{subfigure}{0.5\textwidth}
\includegraphics[width=1\textwidth]{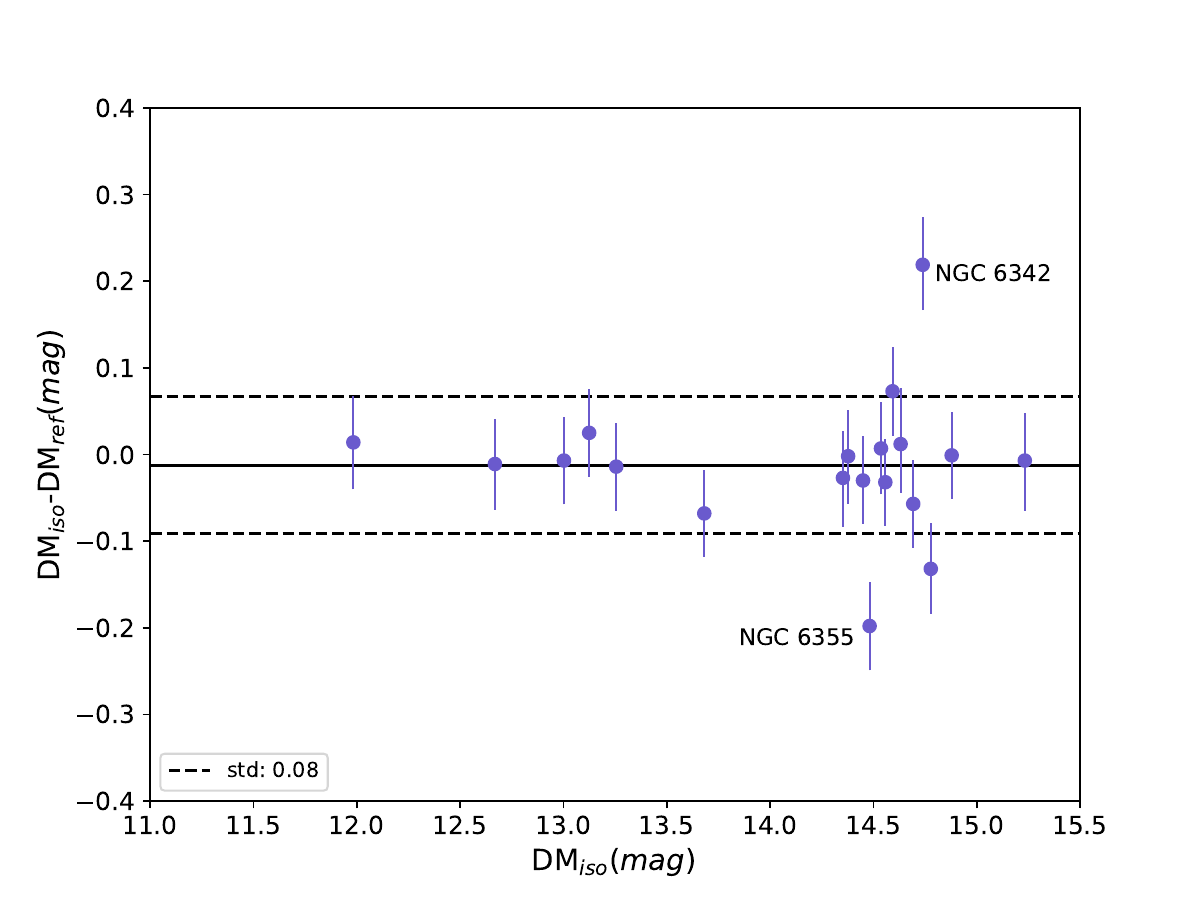} 
\caption{}
\end{subfigure}
\begin{subfigure}{0.5\textwidth}
\includegraphics[width=1\textwidth]{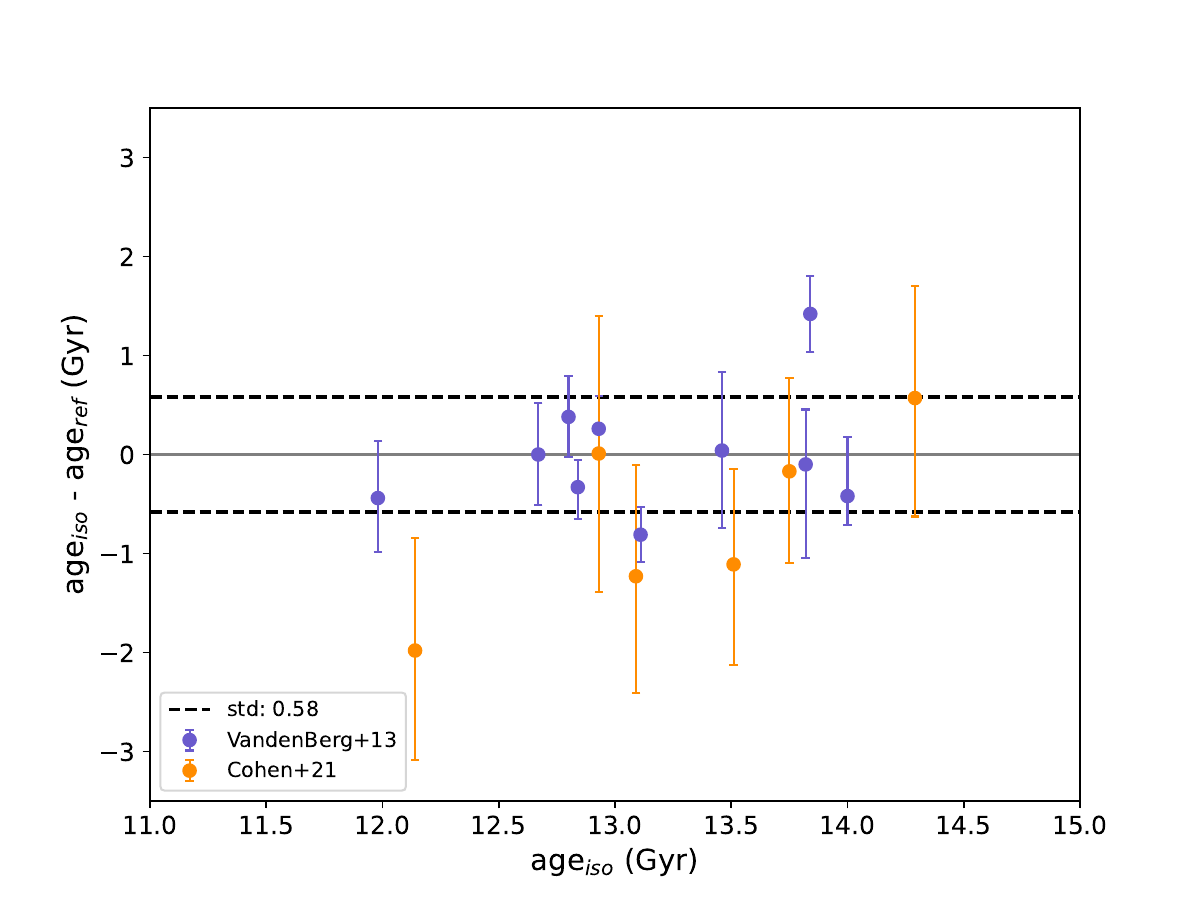} 
\caption{}
\end{subfigure}
\caption{Results from our isochrone fitting procedure, compared to values found in the literature. In panel (a) and (b) we compare our values for [M/H] and E(B-V) respectively with those in the Harris catalogue \citep{harris96}. In panel (c), our resulting distance moduli are compared to the \cite{2021Baumgardt} catalogue. Panel (d) displays our ages in comparison to those in \cite{VandenBerg_2013} (in purple) and in \cite{2021_Cohen} (in orange). We stress out that these are relative age estimates, therefore based on different absolute age scales. To overcome this limit, we absorbed the mean offset between our ages and the literature values so that the mean difference is centred on zero. In all panels, the standard deviations, labelled as "std" are represented as dashed lines.}
\label{fig:comparisons}
\end{figure}

In this section we compare the results from the isochrone fitting procedure with the results available in the literature. These are shown in Fig. \ref{fig:comparisons} for metallicity, colour excess, distance and age, respectively. In each plot, a solid line represents the mean difference between our results and the references, while the dashed lines stand for the associated standard deviation ($1\sigma$ limits). The errorbars were computed summing up in quadrature this standard deviation with the errors associated to each isochrone fit measurement.
We proceed to analyse each plot in further detail:
\begin{itemize}
    \item Metallicity (Panel a): We took as reference literature values those from the Harris catalogue for the majority of our GCs, except for NGC 6642 and NGC 6342. For the former, we adopted as reference the very recent high resolution spectroscopy result from \cite{Geisler_2025} reading [Fe/H]$\simeq$-1.11, perfectly in agreement with our value. For the latter, the only high-resolution spectroscopy measurement available dates back to \cite{2005Origlia}, which we take as our reference in the plot.\\
    \item Colour excess (Panel b): We found all our results fairly in agreement with the Harris catalogue. Indeed, as can be appreciated in the plot, all outputs enter the $1\sigma$ limit within the errors.\\
    \item Distance Modulus (Panel c): Our results are compared with those from \cite{2021Baumgardt}. We first notice that the spread in the difference between our results and the literature generally increases at increasing distance, as expected. The most significant outlier is NGC 6342, which is found to have a distance modulus higher by an amount of $\simeq 0.2$ mag with respect to the literature reference. This GC is moderately reddened and, as previously discussed, lacks recent high resolution metallicity estimates. We therefore infer that the discrepancies between our results and those reported in the literature, both in metallicity and in distance, are primarily attributable to the difficulties inherent in analysing this particularly distant and reddened bulge cluster. This is likely the case for NGC 6355, too, for which our distance modulus appears underestimated with respect to the literature by $\sim$ 0.2 mag.\\
    \item Ages (Panel d): The age comparisons are done purely on a relative scale, and take as reference the relative GC ages derived in \cite{VandenBerg_2013} (purple errorbars), and those from \cite{2021_Cohen} (yellow errorbars). Note that we reported 16 out of our total sample of 18 GCs, since for NGC 6791 and Pal 1 there were no available relative age measurements comparable to ours. The mean offset due to the relative scale is absorbed such that the mean difference is centred on zero, and we can see that in this way only two GCs barely exceed the $1\sigma$ limit, namely NGC 6642 and NGC 6637. These are two bulge GCs, for which the uncertainties in the age measurements can increase due to the already discussed challenges in measuring the metallicity or any possible internal variation in helium or iron abundance, and most importantly, reddening and distance.
\end{itemize}
\section{Robustness of the peak-merger associations} 
\begin{figure}[h!]
    \centering
\includegraphics[width=0.8\linewidth]{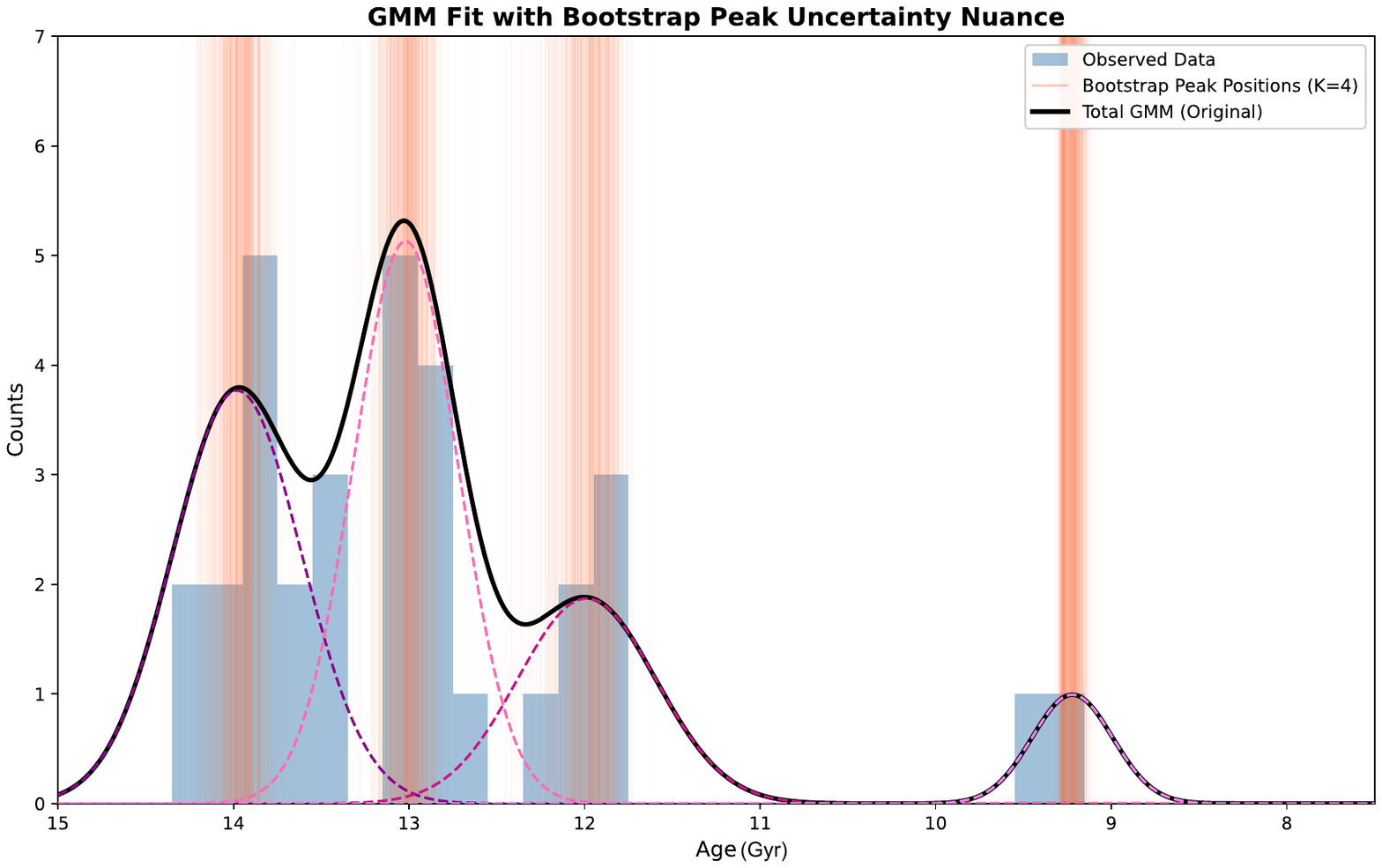}
    \caption{The figure displays the same plot as in Figure \ref{fig:gmm_hist}, but with the superimposition of the means resulting from the first 400 bootstrap realisations (solid, orange vertical lines) having the best number of components according to the BIC (which is 4).}
    \label{fig:bootstrap}
\end{figure}
\label{app:bootstrap}
To verify the robustness of the four peaks of in-situ GC formation we found in Sect. \ref{sec:adf}, we performed a bootstrap test with replacement on the GC sample, iterating 2000 times. The GMM procedure was then applied to each bootstrap realisation, always evaluating a number of components ranging from 1 to 10, and the corresponding BIC was calculated for each run. We found that the most likely case, which is favoured in $\sim$ 47\% of the iterations, is a four-component model. Another relatively frequent case is the one for 5 components, which is found in $\sim$ 25\% of the cases. In this case, though, the oldest peak is split in two components, whereas the other three younger peaks are left unaltered. This means that all the components associated with past merger events are effectively stable in $\sim$72\% of the realizations. None of the other cases is ever favoured in more than $\sim28\%$ of the iterations. Moreover, we computed the mean value and the standard deviation associated with each of the four components, finding in order of look-back times: 13.96$\pm$0.10 Gyr, 13.01$\pm$0.09 Gyr, 11.99$\pm$0.15 Gyr and 9.23$\pm$0.04 Gyr. While these findings demonstrate the stability of the peaks, the quoted errors apply exclusively to the four-component model, and do not include the uncertainties in the number of components.
\newpage
\section{Re-calibrating the age of M4} \label{app:m4}
\begin{figure}[h!]
    \centering
    \includegraphics[width=0.7\linewidth]{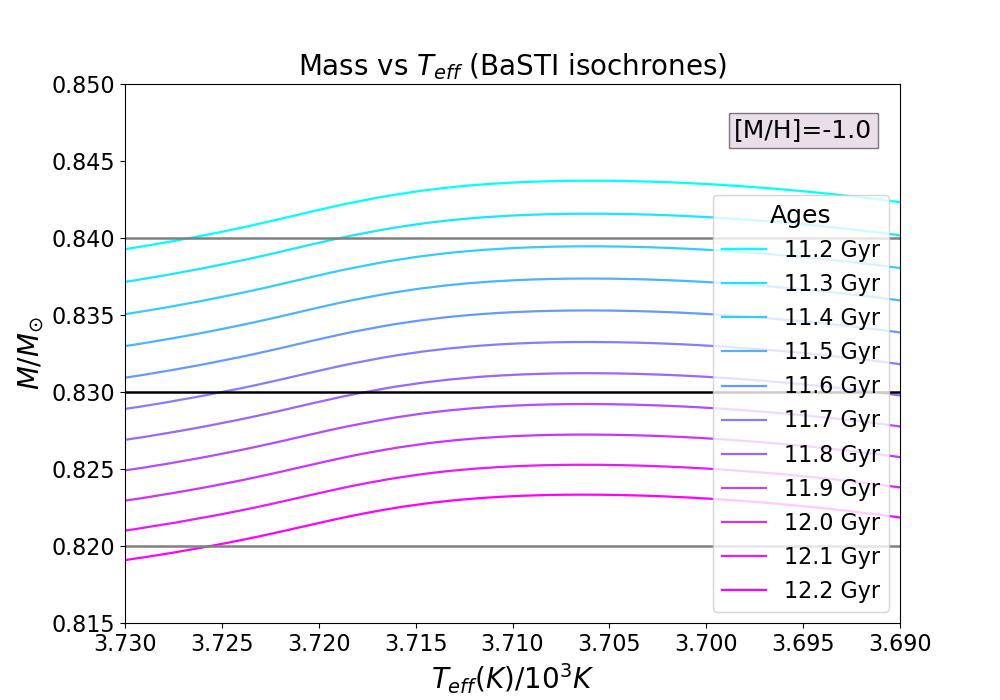}
    \caption{Isochrones at the mean metallicity of [M/H]=-1.0 for M4 in the mass vs effective temperature plane. The black solid line refers to the mean low-RGB mass found in \cite{Howell2022} for this cluster. The gray solid lines represent the associated errors.}
    \label{fig:m4}
\end{figure}

To assess the viability of extending our age-scale re-calibration to the full sample of GCs, we applied the procedure adopted for NGC 6791 to the cluster M4 (also known as NGC 6121). This GC has been analysed in the CARMA framework already, and has been associated to the LHK merger event in M26, with an age of 12.24$^{+0.22}_{-0.30}$ Gyr, and a global metallicity [M/H]=-1.01$^{+0.07}_{-0.05}$. \\
We point out that, for this cluster, only asteroseismic masses derived through scaling relations are available. In particular, \cite{Howell2022} found a mass for low-RGB stars in M4 of 0.83 $\pm 0.1 M_{\odot},$ in perfect agreement with the results in \cite{2016Miglio}, reading 0.84 $\pm 0.1 M_{\odot}.$ These results were obtained using the same scaling relations, and their tight consistency suggests a high accuracy for the adopted scaling relation, as stated in \cite{Howell2022}. Since the catalogue used in this latter analysis has a significantly larger number of stars, we use their mass value to carry on the analysis.\\
Following the same criteria explained in Sect. \ref{sec:absolute}, from the plot in Fig. \ref{fig:m4} we infer a re-calibrated age of 11.7 $\pm 0.5$ Gyr for M4. This leads to a difference of 0.54 between the CARMA age and the newly derived one, consistent with the offset derived for NGC 6791.

\FloatBarrier
\twocolumn

\clearpage

\end{appendix}
\end{document}